\documentclass[useAMS,usenatbib]{mnras}
 \usepackage{times,graphicx,amssymb,amsmath,natbib}
\usepackage[pdftex]{}
\pdfoutput=1
\newcommand{\be}{\begin{equation}}
\newcommand{\e}{\end{equation}}
\newcommand{\bear}{\begin{eqnarray}}
\newcommand{\ear}{\end{eqnarray}}

\newcommand{\de}{{\rm d}}

\def\nat{Nature}

\def\mnras{MNRAS}
\def\apj{ApJ}
\def\apjl{ApJL}
\def\apjs{ApJS}
\def\aap{A\&A}

\def\araa{ARA\&A}

\begin{document}

\title[Equilibrium model prediction for MS scatter]
{Equilibrium model prediction for the scatter in the star-forming main sequence}

\author[Mitra, Dav\'e, Simha \& Finlator]{
\parbox[t]{\textwidth}{\vspace{-1cm}
Sourav Mitra$^{1}$\thanks{E-mail: hisourav@gmail.com},
Romeel Dav\'e$^{1,2,3}$,
Vimal Simha$^{1}$,
Kristian Finlator$^{4}$}
\\
\\$^1$ University of the Western Cape, Bellville, Cape Town 7535, South Africa
\\$^2$ South African Astronomical Observatories, Observatory, Cape Town 7925, South Africa
\\$^3$ African Institute for Mathematical Sciences, Muizenberg, Cape Town 7945, South Africa
\\$^4$ New Mexico State University, Las Cruces, NM, USA
} 

\maketitle

\date{\today}

 \begin{abstract}
The analytic ``equilibrium model" for galaxy evolution using a mass
balance equation is able to reproduce mean observed galaxy scaling
relations between stellar mass, halo mass, star formation rate (SFR)
and metallicity across the majority of cosmic time with a small
number of parameters related to feedback. Here we aim to test this
data-constrained model to quantify deviations from the mean relation
between stellar mass and SFR, i.e. the star-forming galaxy main
sequence (MS). We implement fluctuation in halo accretion rates
parameterised from merger-based simulations, and quantify the
intrinsic scatter introduced into the MS under the assumption that
fluctuations in star formation follow baryonic inflow fluctuations.
We predict the 1-$\sigma$ MS scatter to be $\sim0.2-0.25$ dex over
the stellar mass range $10^8 M_\odot$ to $10^{11} M_\odot$ and a
redshift range $0.5\la z\la 3$ for SFRs averaged over 100~Myr.  The
scatter increases modestly at $z\ga 3$, as well as by averaging
over shorter timescales.  The contribution from merger-induced star
formation is generally small, around 5\% today and $10-15$\% during
the peak epoch of cosmic star formation.  These results are generally
consistent with available observations, suggesting that deviations
from the MS primarily reflect stochasticity in the inflow rate owing
to halo mergers.
\end{abstract}


\section{Introduction}

Over the past few years, advances in large multi-wavelength galaxy
surveys have considerably increased our knowledge of galaxy evolution.
In particular, such surveys have served to greatly constrain the
scaling relations between global galaxy properties, which places
constraints on the nature of galaxy growth.  As observational
uncertainties, both statistical and systematic, are lowered, key
galaxy scaling relations have been shown to be quite tight, such
as the relationship between stellar mass and gas-phase metallicity
that shows a scatter of $\sim 0.1$~dex or less \citep{2004ApJ...613..898T}.
These scaling relations suggest an underlying simplicity involved
in the various complex processes of galaxy evolution.

A particularly well-investigated scaling relation is that between
the star formation rate (SFR) and stellar mass ($M_*$) in star-forming
galaxies (SFGs), colloquially known as the SFG ``main sequence"
(MS).  The MS extends over several orders of magnitudes in $M_*$
and out to high redshifts, with a modest scatter of $\sim0.3$ dex
\citep{2007ApJ...660L..43N,2007A&A...468...33E,2007ApJ...670..156D,2007ApJS..173..267S,
2012ApJ...754L..29W,2015A&A...575A..74S} which includes both intrinsic
scatter and measurement uncertainties.  The existence of such tight
scatter at all observed epochs suggests that most galaxies assembled
their stellar mass fairly steadily rather than predominantly in
starburst episodes, implying that mergers have a sub-dominant
contribution to the global star formation history
\citep{2011ApJ...739L..40R,2012ApJ...747L..31S, 2015A&A...575A..74S}.

Concurrently, cosmological simulations highlighted the fact that
the inflow of gas fueling star formation into galaxies enters
predominantly in smooth, cold
accretion~\citep{2005MNRAS.363....2K,2009Natur.457..451D}.  However,
it has long been recognised that unabated accretion would result
in galaxies far too large compared to
observations~\citep{1991ApJ...379...52W,2001MNRAS.326.1228B}, and
solving this so-called ``overcooling problem" requires strong
feedback to suppress star formation.  The now-ubiquitous observations
of galactic outflows in SFGs over much of cosmic
time~\citep{2005ApJ...621..227M,
2009ApJ...692..187W,2010ApJ...717..289S,2014ApJ...794..156R} suggests
that such outflows are the primary mechanism for self-regulation
in SFGs, by ejecting copious amounts of gas from galaxies that would
otherwise form into stars.  Simulations now routinely include such
outflow processes in order to achieve good agreement with basic
galaxy demographics~\citep{2014MNRAS.437..415D,2014MNRAS.445..175G,
2015MNRAS.454.2691M,2015MNRAS.446..521S,2015ARA&A..53...51S,dav16,
2016ApJ...824...57C,2016ApJ...827L..23W}. Furthermore,
models argue that the return of some outflow material, often called
``wind recycling", is also an important component of inflow required
to match galaxy
properties~\citep[e.g.][]{2010MNRAS.406.2325O,2013MNRAS.431.3373H}. Thus
it appears that, within the well-constrained growth of large-scale
structure, the {\it baryon cycling} processes of inflows, outflows,
and wind recycling govern the growth of galaxies across cosmic time.

While the physical mechanisms and dynamics require complex
cosmologically-based simulations to fully describe, it is possible
to obtain intuitive insights and robust constraints on baryon cycling
processes using a simple analytic framework.  The essential equation
balances the gas inflow rate into the interstellar medium (ISM) of
galaxies, versus the sum of the mass outflow rate and star formation
rate, as well as fluctuations in the gas reservoir.  Such models
are commonly referred to as ``equilibrium"
\citep{2008MNRAS.385.2181F,dav12}, ``gas regulator"
\citep{2013ApJ...772..119L, 2014MNRAS.443.3643P}, or ``bathtub"
\citep{2010ApJ...718.1001B,2014MNRAS.444.2071D} models.
\citet{2008MNRAS.385.2181F} crucially pointed out that simulations
predict that the rate of change of the gas reservoir is small
compared to the other terms, and setting this term exactly to zero
results in simplifications that make the model more intuitive and
insightful, while still being a realistic description of galaxy
growth averaged over cosmological timescales.  We call this assumption
of a non-evolving gas reservoir the ``equilibrium assumption," from
which the equilibrium model follows~\citep{dav12}.  Notably, the
gas regulator model does not make this assumption.  Regardless,
this simple framework is able to capture the essential baryon cycling
processes analytically, thereby enabling a more intuitive view of
how galaxy growth proceeds.

The next step in such models was to constrain the free parameters
associated with baryon cycling.  In \citet{Mitra15}, hereafter
Paper~I, we parameterised the equilibrium model with three variables
corresponding to ejective feedback via a mass loading factor ($\eta$),
preventive feedback via an evolving halo mass scale for quenching
($\zeta$), and wind recycling via a typical recycling time for
ejected material to re-accrete ($t_{\rm rec}$).  For each variable,
we postulated simple dependences on halo mass and redshift, along
with an overall amplitude, resulting in 9 free parameters.  We found
that our null hypothesis of a halo mass quenching scale of $\approx
10^{12}M_\odot$ at $z=0$ was preferred by the Bayesian evidence,
which reduced our number of free parameters by one.  With these 8
free parameters, we then fit to observations of the stellar mass--halo
mass relation, the MS, and the mass-metallicity relation from
$z=0-2$, using a Monte Carlo Markov Chain (MCMC) algorithm.  We
obtained a best-fit reduced $\chi^2=1.6$ to all the data at all
those epochs, which is significantly better than is typically
obtained in simulations or semi-analytic
models~\citep{2015ARA&A..53...51S}.  This demonstrates that the
baryon cycling framework in the equilibrium model can provide a
good description of galaxy growth, and moreover provides meaningful
constraints on the baryon cycling variables themselves.

Our equilibrium model results suggest that one can fit the mean
galaxy scaling relations and their cosmological evolution without
explicitly including mergers.  Nonetheless, mergers add stochasticity
to galaxy evolution that was not accounted for in the \citet{Mitra15}
model.  In that sense, Paper I reflects a {\it first order} model
for galaxy evolution which only accounts for the mean evolution of
the scaling relations.  Meanwhile, the scatter around the mean
scaling relations are driven by other processes such as environment
and the fluctuations in the inflow rate owing to mergers
\citep{dav11a,Mitra15,2015MNRAS.454..637G}.  Such processes thus
can be regarded as yielding {\it second order} deviations from the
mean relations.

In this paper, we provide a quantitative test of the
earlier first order equilibrium model
by investigating the variations around mean trends in
the main sequence.  Our basic aim is to see how the inflow fluctuations,
predicted from a merger-tree based approach, can give rise to
observed scatter in the MS.  Other groups have likewise investigated
this \citep{2010MNRAS.405.1690D,2014MNRAS.443..168F,2015MNRAS.447.3548S,
2016MNRAS.455.2592R}, generally finding $0.1-0.4$ dex scatter in
the stellar mass range $10^9M_\odot$ to $10^{11}M_\odot$, but have
not done so within a MCMC-constrained equilibrium model-type framework
as we do here.

This paper is structured as follows. In the next section, we review
the key features of our basic equilibrium model along with the
modifications made for the purpose of this work, and present the
(minor) updates to the parameter constraints when including inflow
fluctuations. We then present the resulting scatter in MS obtained
from our model and compare it with the present observations in
Section \ref{sec:results}. Finally, we summarize and conclude our
main findings in Section \ref{sec:summary}.

\section{Model Description}\label{sec:model}

\subsection{First order model}\label{subsec:zeroethorder}

We begin by summarizing the main features of the basic equilibrium
model, built on a simple set of equations which well approximates
galaxy evolution in full hydrodynamic simulations
\citep{2008MNRAS.385.2181F}. Unlike SAMs, these models are not based
on halo merger trees, nor do they attempt to track the formation
of a disk and subsequent mergers, as in traditional galaxy formation
theory \citep{1991ApJ...379...52W,1998MNRAS.295..319M}. Instead,
we rely on the view that galaxies grow along a slowly-evolving
equilibrium between accretion, feedback, and star formation as
outlined in
\cite{2008MNRAS.385.2181F,2010ApJ...718.1001B,dav12,2013ApJ...772..119L}:
\begin{equation}\label{eqn:equil}
\dot{M}_{\rm in} = \dot{M}_{\rm out} + {\rm SFR},
\end{equation}
where $\dot{M}_{\rm in}$ is the mass inflow rate onto the galaxy’s
star-forming region and $\dot{M}_{\rm out}$ is outflow rate. This
is a simple mass balance equation with an extra assumption that the
net change of gas mass within the ISM is zero
\citep[i.e. the equilibrium assumption;][]{2008MNRAS.385.2181F,2012ApJ...753...16K,
2013ApJ...768...74T,2013ApJ...778....2S}. From this it is possible to derive the
equations for the star formation rate and metallicity within the
ISM as (see \citealt{dav12} and \citealt{Mitra15} for details):
\begin{equation}\label{eqn:SFR}
{\rm SFR} = \frac{\zeta\dot{M}_{\rm grav}+\dot{M}_{\rm recyc}}{1+\eta},
\end{equation}
and
\begin{equation}\label{eqn:Z}
Z_{\rm ISM} = \frac{y{\rm SFR}}{\zeta\dot{M}_{\rm grav}},
\end{equation}
where $\dot{M}_{\rm grav}$ is the gravitational-driven inflow of dark matter halos that
is an outcome of $\Lambda$CDM cosmology,
$y$ is the metal yield, and $\dot{M}_{\rm recyc}$ is accretion rate of the material that was
previously ejected in outflows. The above relations contain three unknown variables: $\eta$
(mass loading factor or the ejective feedback parameter), $\zeta$ (preventive feedback)
and $t_{\rm rec}$ (wind recycling time or the recycling parameter) which are collectively
known as {\it baryon cycling parameters} \citep{dav12}.

\begin{figure*}
   \includegraphics[height=0.67\textwidth, angle=0]{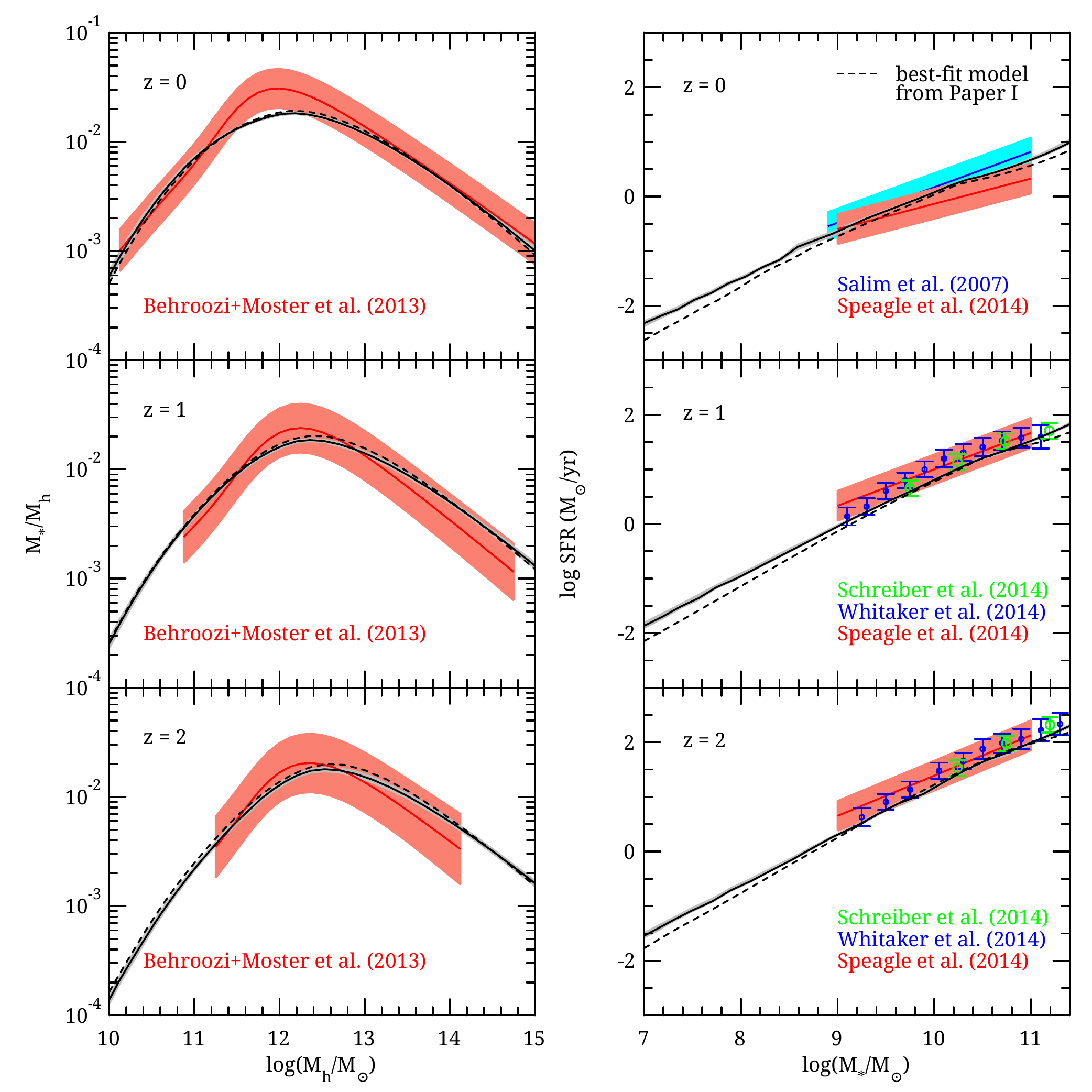}   
\caption{MCMC constraints on our first order equilibrium model for
the stellar mass-halo mass relation ({\it left panel}) and the SFR-$M_*$
relation ({\it right panel}) at $z=0,1,2$. The solid black lines denote the
best-fit model, whereas the thin gray shaded regions refer to their
1-$\sigma$ confidence limits. All the errors on observed data sets,
described in \S\ref{subsec:zeroethorder}, indicate corresponding 1-$\sigma$
uncertainties. The best-fit model from our Paper I is shown by dashed black
lines for comparison.}
\label{fig:MCMC}
\end{figure*}

Despite ongoing efforts and improvements, the baryon cycling parameters remain difficult to
constrain observationally, because inflows and outflows generally occur in
diffuse, multi-phase circum-galactic gas which is challenging to fully
characterise via either absorption or emission probes.  The equilibrium
model, instead, provides a way to constrain these parameters from
the global demographic evolution of the galaxy population, within the context
of the baryon cycling paradigm.  To do
so, we represent them by 8 free variables that quantify their
behavior with halo masses and redshifts \citep{Mitra15}:
\begin{equation}\label{eqn:etaparam}
\eta = \left(\frac{M_h}{10^{\eta_1+\eta_2\sqrt{z}}}\right)^{\eta_3}
\end{equation}
\begin{equation}\label{eqn:trecparam}
t_{\rm rec} = \tau_1\times10^9{\rm yr}\times(1+z)^{\tau_2} \left(\frac{M_h}{10^{12}}\right)^{\tau_3}
\end{equation}
\begin{equation}\label{eqn:zetaparam}
 \zeta_{\rm quench} = {\rm MIN}\left[1,\left(\frac{M_h}{M_q}\right)^{\zeta_1}\right], 
 \frac{M_q}{10^{12} M_\odot} = (0.96 + \zeta_2 z),
\end{equation}
where $\zeta_{\rm quench}$ is quenching feedback parameter and $M_q$ is the
quenching mass. We then employ a Bayesian
MCMC approach using recent measurements of three well-known galaxy
scaling relations that relate the halo mass, stellar mass, SFR and
metallicity of galaxies: (i) the stellar mass ($M_*$) vs. halo mass
($M_h$) (SMHM) relation \citep{2013ApJ...770...57B,2013MNRAS.428.3121M},
(ii) the stellar mass vs. gas-phase metallicity (MZR) relation
\citep{2013ApJ...765..140A,2014ApJ...791..130Z,2014ApJ...795..165S,2015ApJ...799..138S}
and (iii) the stellar mass vs. SFR relation
\citep{2014ApJS..214...15S,2014ApJ...795..104W, 2015A&A...575A..74S}.
To properly represent the evolution of the galaxy population, we
consider these relations over a significant fraction of cosmic time,
at redshifts $z=0$ (today), $1$ ($\sim6$~Gyr ago), and $2$ ($\sim10$~Gyr
ago). We refer the reader to Paper I for a detailed description
of this analysis method. The only modification we make here is that,
now we compute the $\dot{M}_{\rm grav}$ from simulations in a
different way (discussed later in this section), rather than what
we used earlier (i.e. a simple fitting formula from
\citealt{2009Natur.457..451D}).

We show the match between those scaling relations and our best-fit
model predictions in Figure~\ref{fig:MCMC} by solid black lines,
which are quite similar to what we obtained in Figure~2 of Paper
I, shown in dashed lines. Note that, all the error bars here reflect
1-$\sigma$ or 68\% confidence limits (C.L.) around the mean. The
agreement is again quite good, with an overall reduced chi-squared
value of $\approx 2$. The best-fit values and 68\% C.L.  for all 8
parameters are listed in Table~\ref{tab:MCMC-results}.  The results
from Paper I are also shown here for comparison, which are again
comparable to what we obtain here.  We use the Bayesian evidence
to ensure that removing any one of these 8 parameters is not
statistically favored. Overall, neither hydrodynamic simulations
nor SAMs employing many more parameters are able to achieve such a
good match across such a wide a range of redshifts.  Note that up
till now, we have not explicitly considered the scatter around the
scaling relations, and instead we only aim to fit the mean trends.
As such, we refer to this as the {\it first order} equilibrium model
for galaxy evolution. In the next section, we shall see how one can
get a reasonable scatter in $M_*$-SFR relation from the fluctuations
of inflow rates using a simple probabilistic approach.

\begin{table}
\begin{center}
\begin{tabular}{l|c|c}
Parameters & \multicolumn{2}{c}{Best-fit value and 1-$\sigma$ errors}\\
& This paper & Paper I\\
\hline
\hline\\
$\eta_1$ & ~$10.85_{-0.06}^{+0.06}$ & ~$10.98_{-0.10}^{+0.07}$ \\\\
$\eta_2$ & ~~$0.81_{-0.07}^{+0.07}$ & ~~$0.62_{-0.06}^{+0.07}$ \\\\
$\eta_3$ & $-1.15_{-0.07}^{+0.08}$ & $-1.16_{-0.06}^{+0.06}$ \\\\
$\tau_1$ & ~~$1.12_{-0.24}^{+0.37}$ & ~~$0.52_{-0.07}^{+0.24}$ \\\\
$\tau_2$ & $-0.62_{-0.21}^{+0.11}$ & $-0.32_{-0.20}^{+0.06}$ \\\\
$\tau_3$ & $-0.47_{-0.06}^{+0.05}$ & $-0.45_{-0.07}^{+0.10}$ \\\\
$\zeta_1$ & $-0.45_{-0.07}^{+0.07}$ & $-0.49_{-0.08}^{+0.07}$  \\\\
$\zeta_2$ & ~~$0.51_{-0.17}^{+0.15}$ & ~~$0.48_{-0.12}^{+0.13}$ \\\\
\hline
\end{tabular}
\caption{MCMC results from first order model: best-fit values and 68\% confidence limits
on the all eight parameters for this paper and our Paper I.} 
\label{tab:MCMC-results}
\end{center}
\end{table}

\subsection{New features}\label{subsec:newfeatures}

We now discuss the additional features of our model which we have
implemented in this work to generate the scatter in halo accretion
rate.  To do so, we now need to consider the fact that the accretion
of material into halos is not smooth, but rather arrives in lumps.
For the dark matter, this corresponds to halo merging, which we can
express using an analytical fitting formula for the dimensionless
mean merger rate, $\de N_m/\de{\epsilon}\de z$, where $\epsilon$
is the merger mass ratio $M_{\rm subhalo}/M_{\rm parent}$
\citep{2010MNRAS.406.2267F}:
\begin{equation}
\frac{\de N_m}{{\de\epsilon}~{\de z}} (M, \epsilon, z) = A \left(\frac{M}{10^{12}M_{\odot}}\right)^{\alpha}
\epsilon^{\beta} \exp\left(\frac{\epsilon}{\bar{\epsilon}}\right)^{\gamma} (1+z)^{\eta}.
\label{mergerrateeqn}
\end{equation}
The free parameters $A$, $\bar{\epsilon}$, $\alpha$, $\beta$, $\gamma$ and $\eta$
are obtained by fitting to an N-body simulation based on Millennium
\citep{2005Natur.435..629S} and Millennium-II \citep{2009MNRAS.398.1150B}
simulation, but with a cosmology consistent with \cite{2015arXiv150201589P}.
The best-fit values are $\left(A, \bar{\epsilon}, \alpha, \beta, \gamma, \eta\right)
=\left(0.0104, 0.00972, 0.133, -1.995, 0.263, 0.0993\right)$.

The cumulative number of mergers received by a halo of mass $M$,  of objects with mass between
$M\epsilon_{\rm min}$ and $M\epsilon_{\rm max}$, between redshift $z_0$ and $z$, is then given by:
\begin{equation}
N_m = \int_{z_0}^{z} \de z \int_{\epsilon_{\rm min}}^{\epsilon_{\rm max}} \de{\epsilon}
\frac{\de N_m}{{\de\epsilon}~{\de z}} (M, \epsilon, z),
\label{eq:integralN}
\end{equation}
and 
\begin{equation}
\frac{dM}{dz} = \int_{\epsilon_{\rm min}}^{\epsilon_{\rm max}} M~\epsilon~\de{\epsilon}
\frac{\de N_m}{{\de\epsilon}} (M, \epsilon, z),
\label{eq:integraldN}
\end{equation}
from which the mass accretion rate $\dot{M}_{h}$ can be calculated
by a transformation of variables from redshift $z$ to time $t$.
This allows us to compute the baryonic inflow rate as $\dot{M}_{\rm grav}=f_b \dot{M}_h$,
We choose $\epsilon_{\rm min}=10^4M_{\odot}/M_{\rm parent}$,
thereby capturing all mergers down to $M_h=10^4M_\odot$.
We have tested our model with different limits of $\epsilon_{\rm min}$
and although the ``smooth'' component (discussed below) changes a bit,
the overall results remain the same.

For a halo of mass $M$, we compute the number of mergers received
in some time interval $dt$ in bins of mass ratio $\epsilon$
using equation \ref{eq:integralN}. We then sample the distribution
of $N_m$ as a function of $\epsilon$ to obtain the masses of
individual halos that merge with a given halo of mass $M$. Each
such sampling produces a discrete set of masses of halos that merge
with a given parent halo of mass $M$. In details, we split the
merger ratio integral of Equation \ref{eq:integralN} into two
components: (i) {\it Merger component} - first we compute the probability
($P_i$) of a merger in some mass bins by summing up the integrand
of Equation \ref{eq:integralN}.  If $P_i<1$, we generate a uniform
random number between $0-1$ and accept each merger with that
probability only when $P_i$ is greater than that random number. We
stop this calculation once we get to the regime where the probability
is greater than one. (ii) {\it Smooth component} - we do the normal
integration down to $\epsilon_{\rm min}$ from the point where we
stopped the previous calculation. Finally, adding up both components
will give the total $\dot{M}_{h}$. We follow the same procedure for
many different realizations and obtain an array of different accretion
rates for a given $M$.  In this manner, we obtain the the inflow
rate including fluctuations owing to halo mergers.

We have implemented this into the equilibrium model and done an
MCMC fit the mean scaling relations as was done in Paper~I.  This
results in the fits previously shown in Figure~\ref{fig:MCMC}, with
the parameter constraints listed in Table~\ref{tab:MCMC-results}.
The fits are very similar to those arising from the first-order
model.

\section{Results and Discussion}\label{sec:results}

We now turn our attention to the main aim of this paper, namely to
understand the origin of the scatter in the star-forming main
sequence.  We will assume that the fluctuations in the halo accretion
rate are reflected directly in fluctuations in the star formation
rate, as per equation~\ref{eqn:SFR}.  While it clearly takes some
time for the halo inflow to reach the galaxy and form stars, we are
assuming that once equilibrium is reached, the statistical variations
are the same.  To examine this in more detail, we begin by quantifying
the fluctuations in $\dot{M}_h$.

\subsection{Halo inflow rate fluctuations}

\begin{figure*}
   \includegraphics[height=0.38\textwidth, angle=0]{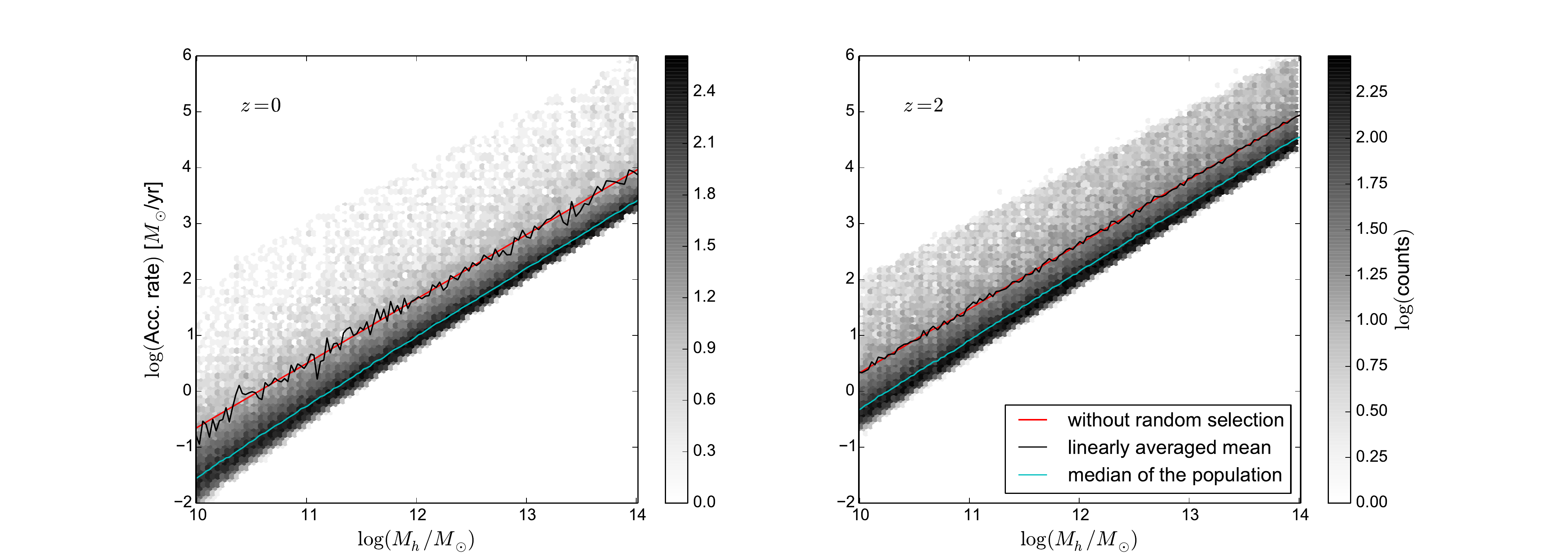}   
\caption{{\it Total} fluctuations in halo mass accretion rates at
redshifts $z=0$ and $2$ for $1000$ different realizations. The
black and cyan curve represent respectively the linearly averaged
mean and median of that distribution, whereas the red one denotes the
accretion rates of the first order model.}
\label{fig:scatterMhdot}
\end{figure*}

Figure~\ref{fig:scatterMhdot} displays the {\it total} variations
in the accretion rate ($\dot{M}_h$) as a function of halo mass at
$z=0$ ({\it left panel}) and $z=2$ ({\it right panel}). At a fixed
halo mass, we get a range of accretion rates, denoted by gray
points, owing to our random selection approach over 1000 trials. The average accretion
rate in bins of $M_h$ is denoted by the solid black line, which can be
compared to the red line which is the smooth halo mass accretion
rate we used in Paper~I.  On average, the halo accretion rate is
well-represented by sampling the stochastic distribution.  

For $z=2$, the black curve is a very close match to the the first-order
mean inflow rate, as most of growth is predominantly in the ``smooth''
mode (i.e. probability $P_i\gtrsim1$).  Towards lower redshifts,
the merger contribution from stochasticity becomes more significant,
which is consistent with trends seen from hydrodynamic
simulations~\citep[e.g.][]{2005MNRAS.363....2K}.
The {\it median} of the population is shown here by the solid cyan
line which lies towards the denser ends of $\dot{M}_h$-distribution as this
quantity is less affected by extreme outliers than the mean\footnote{
The reason for choosing mean over the median in our MCMC analysis
is that the former quantity is more easy (and faster) to calculate.
Also the mean (black curve) closely matches the smooth halo accretion
rates from the simulation (without random selection, red curves),
whereas median underestimates those. So these extreme outliers are
also significant to capture the whole picture.}.

Here we are interested in the scatter about this mean relation.
The scatter appears to be fairly asymmetric, with a large tail to
higher accretion rate owing to high mass ratio mergers.  A more
quantitative estimate of the scatter can be obtained by fitting a
Gaussian or a double-Gaussian~\citep{2015MNRAS.454..637G} to the
histogram of that distribution; we will investigate this in the
next section.  We find that, although our resulting distribution
of inflow rate seems skewed with a large tail, a single Gaussian
fit turns out to be sufficient for our current purposes.  The tail
of the distribution corresponds to starbursts, which we will quantify
later.

\subsection{The scatter around the main sequence}

To investigate how the scatter in $M_*$-SFR relation arises from
stochasticity in the accretion rate, recall Equation~\ref{eqn:SFR}
which directly relates the SFRs with the halo mass accretion rates
($\dot{M}_{\rm grav}$).  For this paper, we only investigate the
scatter associated with halo inflow.  It is also possible that there
is scatter associated with some of the baryon cycling parameters;
for instance the mass loading factor and preventive feedback may
vary between galaxies at a fixed mass \citep{2014MNRAS.443..168F}.
For simplicity, we do not consider these additional sources of
scatter into our current analysis, rather we restrict ourselves to
determine the main sequence scatter arising only from the dispersion
in inflow rates.  We further note that we are intrinsically making
the assumption that inflow into the halo is instantaneously reflected
in inflow into the ISM; this is of course not true, but modulo a
delay related to the infall time through the halo, the spectrum
of halo inflow fluctuations should generally reflect the ISM inflow
fluctuations.  Finally, here we have estimated the SFR averaged
over $\de t = 100$ Myr, denoted SFR($100$); we will consider other
timescales in \S\ref{subsec:timescale_variability}.

\begin{figure*}
   \includegraphics[height=0.38\textwidth, angle=0]{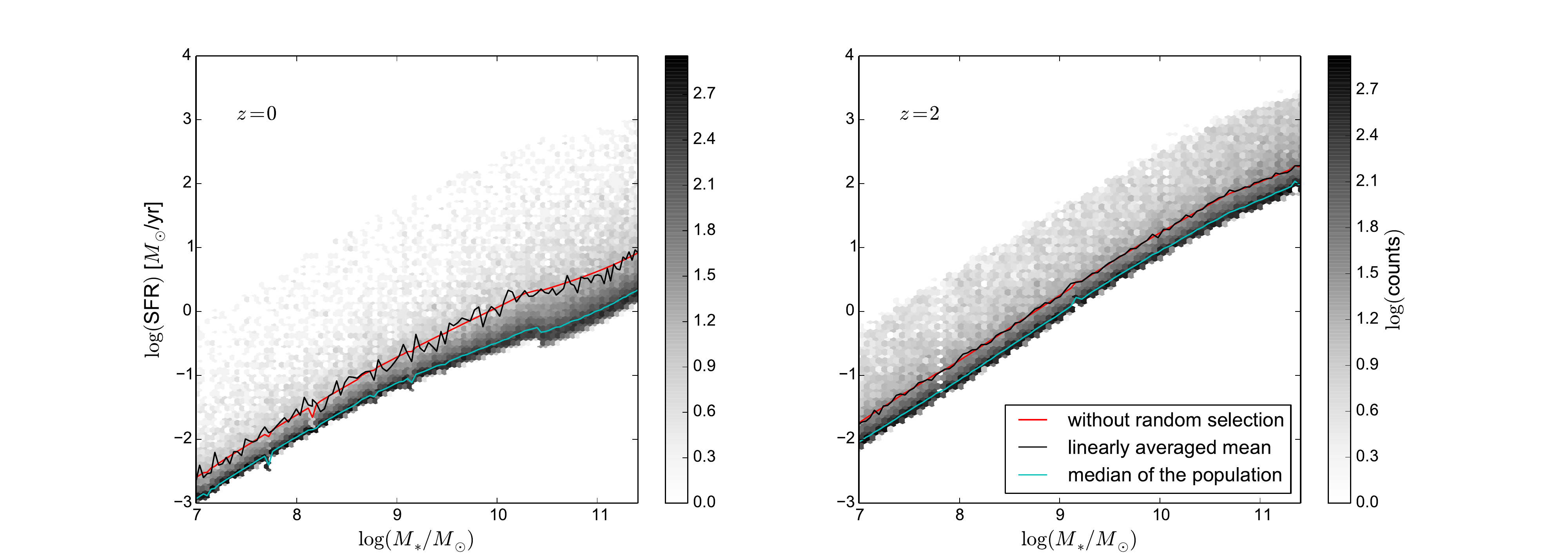}   
\caption{Same as Figure~\ref{fig:scatterMhdot}, but for the $M_*$-SFR relation.}
\label{fig:scatterSFR}
\end{figure*}

The {\it total} variation of the star-forming main sequence is shown
in Figure~\ref{fig:scatterSFR}. Again the linearly averaged mean
agrees with the first order model, as expected due to large number
of samples. The evolution of MS shows a shallower slope at high
stellar masses at later epoch as an outcome of the slowly-decreasing
quenching mass in our model \citep{2015MNRAS.447..374G,Mitra15}.
This behavior is noted in observations as well \citep{2014ApJ...795..104W}.

\begin{figure}
   \includegraphics[height=0.42\textwidth, angle=0]{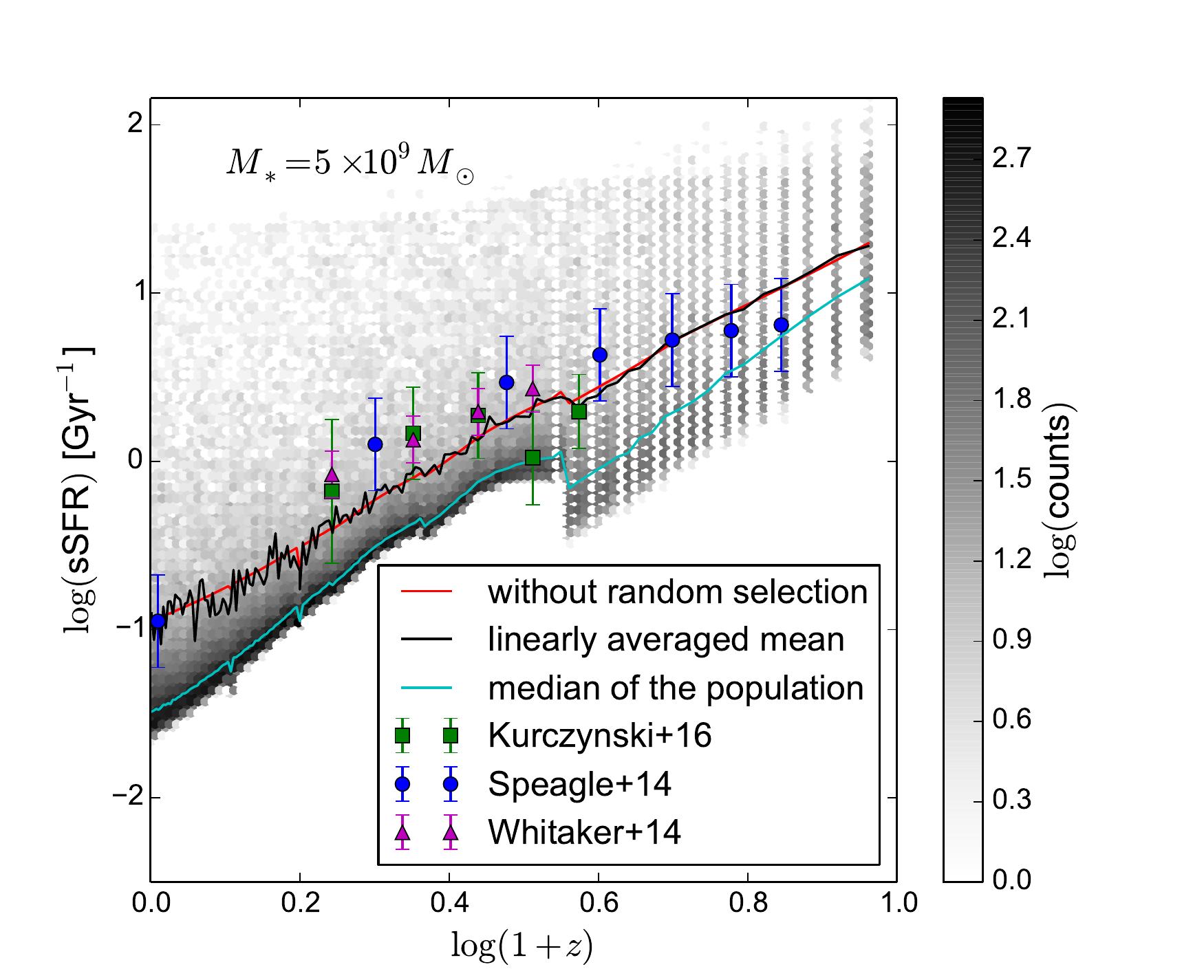}   
\caption{{\it Total} fluctuations in specific star formation rate as function
of redshift at $M_*=5\times10^{9}M_\odot$. The absence of the lumpiness in
linearly averaged mean curves (black) at earlier epoch corresponds to the
halo growth in ``smooth'' mode.}
\label{fig:scattersSFR}
\end{figure}

Additionally, we have also plotted some example star formation histories,
specific star formation rate sSFR($z$), showing the total scatter at
$M_*=5\times10^{9}M_\odot$ in Figure~\ref{fig:scattersSFR}. The sSFR($z$)
is seen to be evolving strongly with redshifts following ${\rm sSFR}\propto(1+z)^b$,
which is again consistent with the observations
\citep{2014ApJ...795..104W,2015MNRAS.453.2540J}. Unsurprisingly, the linearly
averaged mean (black curves) is found to be ``bumpy'' at lower $z$ and then
starts to become smooth and matches the red one at higher redshifts $z\gtrsim2$.
The observed data points shown here are from \cite{2014ApJS..214...15S,
2014ApJ...795..104W,2016ApJ...820L...1K}. Note that, our model seems to
underpredict the sSFR slightly at $z\approx1-2$, which was also seen in
Figure~\ref{fig:MCMC}, but overall is in good agreement with the data
within their observational uncertainties.

To get an estimated scatter in MS, one must fit the probability
distribution function (PDF) of the SFR at some stellar mass and
redshift. We have obtained the number density distribution of
galaxies in six stellar mass bins spanning from $10^8M_\odot$ to
$10^{11}M_\odot$ as a function of their star formation rates at a
range of redshift $0.5\leqslant z\leqslant4$.  We fit Gaussians to
these distributions to objectively identify the MS and its outliers,
similar to the analysis done by \cite{2011ApJ...739L..40R} (but
also see \citealt{2012ApJ...747L..31S} for a double- Gaussian fit).

\begin{figure*}
  \includegraphics[height=0.64\textwidth, angle=0]{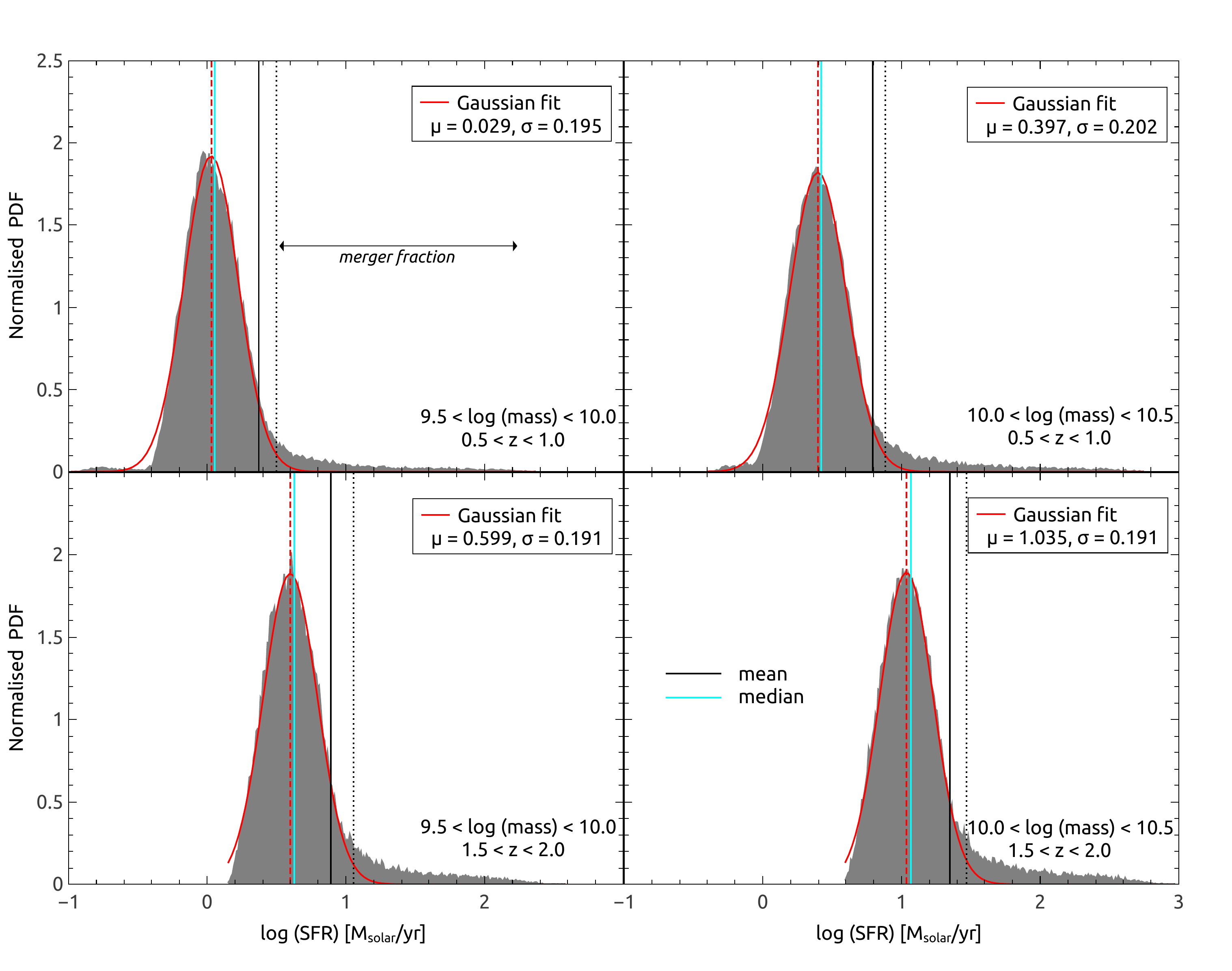}   
\caption{Normalised distributions (solid gray histograms) of SFRs in selected mass-redshift
bins quoted in the figure. Shown in solid red curves are the individual Gaussian fits with a dispersion
of $\sim 0.2$ dex to that distribution. Deviations (dotted black vertical lines) from the Gaussian
are seen at $\sim0.45-0.5$ dex (or $2.4\sigma$) above the mean of the fits (dashed red vertical lines).
The mean and median of those distributions are also shown here by solid black and cyan vertical
lines respectively. Medians are found to be less affected by the extreme outliers at higher SFR ends
and closely match the mode of the distributions, as expected.}
\label{fig:PDF}
\end{figure*}

Figure~\ref{fig:PDF} shows some typical distributions at redshift
range $0.5<z<1$ and $1.5<z<2$, where we plotted the normalised PDFs of the
logarithmic SFR for mass bins $9.5<\log(M_*/M_\odot)<10$ ({\it left
panels}) and $10<\log(M_*/M_\odot)<10.5$ ({\it right panels}).  The
Gaussian fits with the standard deviations $\sigma\approx0.2$ dex
are displayed by the solid red curve.  This log-normal fit in SFR
is quite good over the bulk of the curve, but there is clearly
an excess at high SFR; we will return to this in \S\ref{sec:starbursts}.

\subsection{Comparison to previous observations and models}\label{subsec:previous_works}

A key aim of this paper is to determine whether the simple equilibrium
model can account for the observed scatter in the main sequence
based on fluctuations in the inflow rate.  To this end, we now
compare our scatter predictions, along with their trends with galaxy
mass and redshift, versus a range of recent observations.  We also
discuss comparisons to recent models and simulations that have
predicted the MS scatter.

\begin{figure*}
  \includegraphics[width=0.67\textwidth,keepaspectratio]{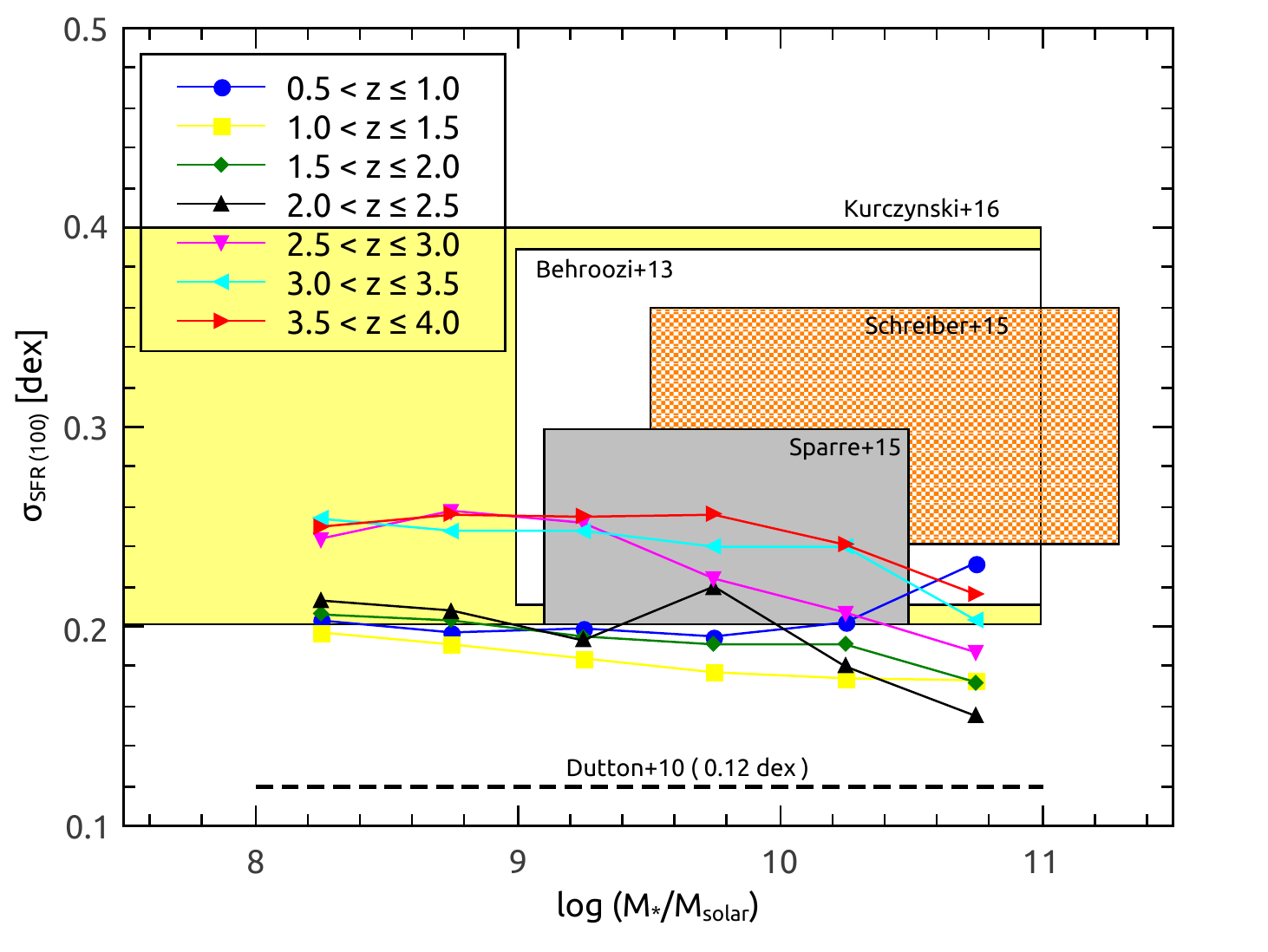}   
\caption{The main sequence scatter (1-$\sigma$) at redshift range $0.5\leqslant z\leqslant4$.
In all cases, we estimate the SFR averaged over $100$ Myr. For comparison, we show the typical 
scatter obtained from observations by \citet{2015A&A...575A..74S} (shaded orange bands)
and \citet{2016ApJ...820L...1K} (shaded yellow; intrinsic scatter of $0.2-0.4$ dex).
For models, we show the Illustris simulation results \citep{2015MNRAS.447.3548S} and results
from the abundance matching models of \citet[see their Table~9]{2013ApJ...770...57B} by the
shaded gray and empty middle bands respectively. The prediction from the analytic disk model of
\citet{2010MNRAS.405.1690D} is shown as the dashed black line.}
\label{fig:sigma}
\end{figure*}

\begin{table*}
\begin{tabular}{c|c|c|c|c|c|c}
redshift range & \multicolumn{6}{c}{$\sigma_{\rm SFR (100)}$ [$\sigma_{\rm SFR (30)}$] in dex}\\
& $8<\log M_*<8.5$ & $8.5<\log M_*<9$ & $9<\log M_*<9.5$ & $9.5<\log M_*<10$ & $10<\log M_*<10.5$ & $10.5<\log M_*<11$\\
\hline
\hline
$0.5\leqslant z\leqslant1$ & $0.203$ [$0.205$] & $0.197$ [$0.202$] & $0.199$ [$0.193$] & $0.195$ [$0.199$] & $0.202$ [$0.199$] & $0.232$ [$0.266$]\\
$1<z\leqslant1.5$ & $0.197$ [$0.216$] & $0.191$ [$0.218$] & $0.184$ [$0.223$] & $0.177$ [$0.211$] & $0.174$ [$0.204$] & $0.173$ [$0.201$]\\
$1.5<z\leqslant2$ & $0.206$ [$0.226$] & $0.203$ [$0.222$] & $0.195$ [$0.216$] & $0.191$ [$0.225$] & $0.191$ [$0.227$] & $0.172$ [$0.207$]\\
$2<z\leqslant2.5$ & $0.213$ [$0.290$] & $0.208$ [$0.277$] & $0.193$ [$0.289$] & $0.220$ [$0.220$] & $0.180$ [$0.210$] & $0.155$ [$0.186$]\\
$2.5<z\leqslant3$ & $0.244$ [$0.251$] & $0.258$ [$0.296$] & $0.252$ [$0.294$] & $0.224$ [$0.305$] & $0.207$ [$0.311$] & $0.187$ [$0.275$]\\
$3<z\leqslant3.5$ & $0.254$ [$0.313$] & $0.248$ [$0.305$] & $0.248$ [$0.307$] & $0.240$ [$0.304$] & $0.240$ [$0.301$] & $0.203$ [$0.282$]\\
$3.5<z\leqslant4$ & $0.250$ [$0.285$] & $0.256$ [$0.317$] & $0.255$ [$0.308$] & $0.256$ [$0.298$] & $0.241$ [$0.285$] & $0.216$ [$0.296$]\\
\hline
\end{tabular}
\caption{Intrinsic scatter around the main sequence of galaxies in bins of different stellar masses and redshifts
averaged over timescale $100$ Myr [$30$ Myr]. Typically, the scatter increases when we decrease the timescale.
Overall, the 1-$\sigma$ dispersion of $\sim 0.2-0.25$ dex for the $100$ Myr case (or $\sim 0.2-0.3$ dex for $30$ Myr)
is broadly comparable with various recent observations.}
\label{tab:sigma}
\end{table*}

Figure~\ref{fig:sigma} shows the estimated intrinsic scatters on
the main sequence for all six stellar mass bins at various redshifts,
computed as the width in dex of the log-normal fit.  The individual
values are presented in Table~\ref{tab:sigma}.  Overall, we find
that the standard deviation $\sigma_{\rm SFR (100)}$ is $\sim
0.2-0.25$ dex, showing no significant trend with redshifts or stellar
mass, except for a weak overall increase at higher redshifts.  

Using the Spitzer MIPS observations, \cite{2007ApJ...660L..43N} and
\cite{2007A&A...468...33E} obtained a $0.3$ dex scatter (1-$\sigma$)
around the MS at $z\sim1$, while \cite{2012ApJ...754L..29W} reported
a dispersion of $0.34$ dex in the range $0<z<2.5$ using a sample
of galaxies selected from the NEWFIRM Medium-Band Survey.  Similarly,
\cite{2011ApJ...739L..40R} determined a value of $0.24$ dex scatter
using mostly UV-derived SFRs. Recently, using the deep UV to NIR
observations in the CANDELS fields, \cite{2015A&A...575A..74S}
reported a scatter around the average SFR to be $\sim0.3$ dex.
These observational results refer to the total scatter, which
includes observational measurement uncertainties.  Several groups
have attempted to correct for measurement errors to determine the
intrinsic scatter.  \cite{2013ApJ...778...23G} found intrinsic sSFR
dispersions of $0.18-0.31$ dex in the stellar mass range of
$9.5<\log(M_*/M_\odot)<11.5$ at $z\sim0.7$, while
\cite{2016ApJ...820L...1K} found an intrinsic scatter of $0.2$
to $0.4$ dex in the redshift range $0.5<z<3$ and in the mass range
$7<\log(M_*/M_\odot)<11$.
Using the MOSDEF Survey of star-forming galaxies with H$\alpha$
and H$\beta$ spectroscopy, \cite{2015ApJ...815...98S} studied the
MS relation at $z\sim2$ and found an intrinsic scatter of $\sim0.31$
dex for the SFR(H$\alpha$) sample, which is $0.05$ dex larger than
what they measured from UV SFRs.
\cite{2014ApJS..214...15S} combined various measurements of the
star-forming MS from literature by recalibrating them to use a common
set of assumptions. After accounting for intrinsic scatter among
SFR indicators, they found that the ``true intrinsic'' scatter is
actually $\sim 0.2$ dex rather than the often reported $0.3$ dex value
and it remains roughly constant over cosmic time.
Generally, observations tend to suggest a
total scatter of $0.3-0.4$~dex depending on mass and redshift, with
the intrinsic scatter being as low as 0.2~dex.

The scatter we predict from the equilibrium model is overall in
very good agreement with these observations.  The model yields an
intrinsic scatter of $\sim 0.2$~dex, with only a very slight mass
dependence increasing to lower masses.  We also don't find a strong
redshift dependence out to $z\sim 2$, though it increases at all
masses at $z\ga 2.5$.

There are indications, particularly from \citep{2016ApJ...820L...1K},
that the intrinsic scatter may be significantly higher than 0.2~dex
in some cases.  Remember that our current approach only includes
scatter associated with inflow fluctuations, while in principle
there could be intrinsic variations in the mass outflow rate ($\eta$)
or in preventive feedback ($\zeta$), which would give extra scatter
to the overall $\sigma_{\rm SFR}$.  Nonetheless, it is interesting
that the lowest measured intrinsic scatters are quite consistent
with our prediction just from inflow fluctuations.  This limits the
stochasticity in outflows to relatively modest values, and suggests
that galactic outflows must be a fairly steady phenomenon at least
when averaged over $\sim 100$~Myr.  Alternatively, there may be
some relationship between outflow rates and inflow stochasticity
such that a relatively tight correlation is maintained owing to
correlated scatter.

Numerous galaxy evolution simulations and semi-analytical or
analytical models have likewise made predictions for the intrinsic
scatter around the $M_*$-SFR relation.  \cite{2015MNRAS.447.3548S}
used the Illustris simulation to obtain a $\sim0.2-0.25$ dex scatter
for $M_*\la 10^{10.5}M_\odot$, while the semi-analytic model for
disk galaxy evolution by \cite{2010MNRAS.405.1690D} predicted a
lower scatter of $0.12$ dex in SFR.  Note that, this value is
much less than the observed one as the later is likely to be dominated
by observational uncertainties. However, this might also reflect the
fact that \cite{2010MNRAS.405.1690D} underestimated the true scatter due to their
simplified treatment of the halo mass accretion history and thus additional
sources of intrinsic scatter are probably required in their model.
Using zoom-in hydro-cosmological simulations of massive galaxies
at $z>1$, \cite{2016MNRAS.457.2790T} examined the evolution of SFGs across
the MS through gas {\it compaction}, {\it depletion} and {\it replenishment}
and measured a {\it true} scatter of $\sim 0.27$ dex with a slightly
increasing trend towards lower redshifts. Note that, if these processes
are independent of inflow fluctuations, then these would contribute
additional scatter around the MS.
Recently, \cite{2016MNRAS.455.2592R}
has reported a $\sigma$ to be $\sim 0.35-0.45$ dex from a simple
analytical approach based on the crucial assumption that the
stellar-to-halo mass ratio is nearly independent of redshift up to
$z\sim4$.  In general, our numbers are similar to that obtained
from Illustris, which supports the notion that MS fluctuations in
hydrodynamic simulations are arising from inflow stochasticity.
The analytic models, on the other hand, have more widely varying
predictions; this may owe to the fact that some of their assumptions
may not be reflective of how galaxy formation proceeds via baryon
cycle-driven growth.

Our method is quite similar in many ways to that in
\cite{2014MNRAS.443..168F}, who examine the scatter in the main
sequence as well as in the MZR and FMR arising from the scatter in
dark matter accretion rates as well as some of the baryon cycling
parameters from their bathtub model. Considering a typical N-body
predicted stochastic scatter in the accretion rates, they found
that the scatter in both the MS and MZR at fixed stellar mass is
comparable to or larger than the observed ones and roughly independent
of halo mass and redshift.  Although we compute stochasticity in
a different way and use a full MCMC approach to characterise the
best-fit relations, 
we echo their general conclusion
that fluctuations in feedback parameters must be sub-dominant.

\subsection{Mass and redshift dependence of the scatter}

The mass and redshift dependence of the MS scatter has also been a
subject of some debate.  Cosmological models like Illustris generally
predict a constant scatter to low masses \citep{2015MNRAS.447.3548S},
although this simulation predicts a growing scatter at high masses
as quenched galaxies enter into their sample.  In contrast, the
FIRE simulations analysed by \citet{2015arXiv151003869S} finds that
$M_*\sim 10^9M_\odot$ galaxies can have extremely bursty SFHs, with
a scatter well over 0.5~dex on short timescales, and argued this
owed to their high resolution and more self-consistent implementation
of feedback processes.

Observations are also starting to characterise the mass dependence
of the MS scatter.  Results from from \cite{2011ApJ...739L..40R},
\cite{2012ApJ...754L..29W}, and \cite{2015A&A...575A..74S}, generally
found the scatter to be independent of $M_*$, and also $z$, down
to $M_*$ as low as $\sim 10^{9.5}M_\odot$ at various epochs.  Also,
\cite{2016MNRAS.455.2592R} did not notice a clear trend with mass
to even lower masses.

In the equilibrium model, the scatter arises purely from halo inflow
fluctuations.  Since halo mass growth rates are fairly self-similar,
one doesn't expect a strong trend with halo mass, and thus stellar
mass, in the scatter.  Hence it is expected that our scatter has
only a very weak mass dependence, which is what we see in
Figure~\ref{fig:sigma}.  This seems to be broadly in agreement with
observations, which again supports the notion that SFR fluctuations
are primarily driven by inflow fluctuations.  In the case of FIRE,
the larger scatter is likely driven by the strong variations in
outflow strength on small timescales in dwarf galaxies; as observations 
of such galaxies improves, this will provide a significant constraint
how bursty low-mass galaxy SFHs can be.

We also find no redshift dependence in the scatter up to $z\sim 2$,
and then a modest increase at higher redshifts.  This is in agreement
with observations by \cite{2016ApJ...820L...1K} who also found
essentially no evolution in intrinsic scatter, with a typical value
of $\sigma\sim 0.25$ from $z\sim 1-3$.  They did find a higher
scatter at $z\la 1$, which likely owes to the inclusion of galaxies
on their way to being quenched.  Meanwhile, our results are in
agreement with the model by \cite{2016MNRAS.455.2592R} who averaged
the halo accretion rate over the dynamical timescale and found no
redshift dependence.  Overall, the predicted lack of a strong redshift
dependence seems consistent with available data.

In summary, the equilibrium model predicts SFR scatter at a given
$M_*$ that is comparable to the lowest values for the observed
intrinsic scatter, and are generally below observed values for the
total scatter.  Models show a larger range in scatter depending on
assumptions and techniques, but the equilibrium model predictions
generally agrees most closely with predictions taken directly from
cosmological hydrodynamic simulations.  We do not find a strong
mass or redshift dependence in the scatter, suggesting that dwarf/early
galaxy SFHs are not expected to be significantly burstier.  This
is generally in agreement with available data, and bolsters our
claim that halo inflow fluctuations, which are expected to be fairly
self-similar with mass, drive SFR fluctuations.

\subsection{Merger-driven starbursts}\label{sec:starbursts}

Not all galaxies lie within the scatter of the main sequence.
Galaxies fall below the main sequence as they are quenched by a
variety of processes such as black hole feedback or gas stripping
processes~\citep{2015ARA&A..53...51S}.  The outliers above the main
sequence are starburst galaxies, and can be identified in our model
as high upwards fluctuations in the inflow rate.  Even though the
equilibrium model does not account for internal dynamical processes
that can drive the most extreme
starbursts~\citep{1996ApJ...464..641M,1996ARA&A..34..749S}, such
galaxies are very rare and require special initial conditions, while
more typical mergers are far less extreme~\citep{2008MNRAS.384..386C}.
We can thus quantify in our model the fraction of galaxies in such
mergers, as well as their global contribution to overall star
formation.

\begin{figure}
  \includegraphics[width=0.5\textwidth,keepaspectratio]{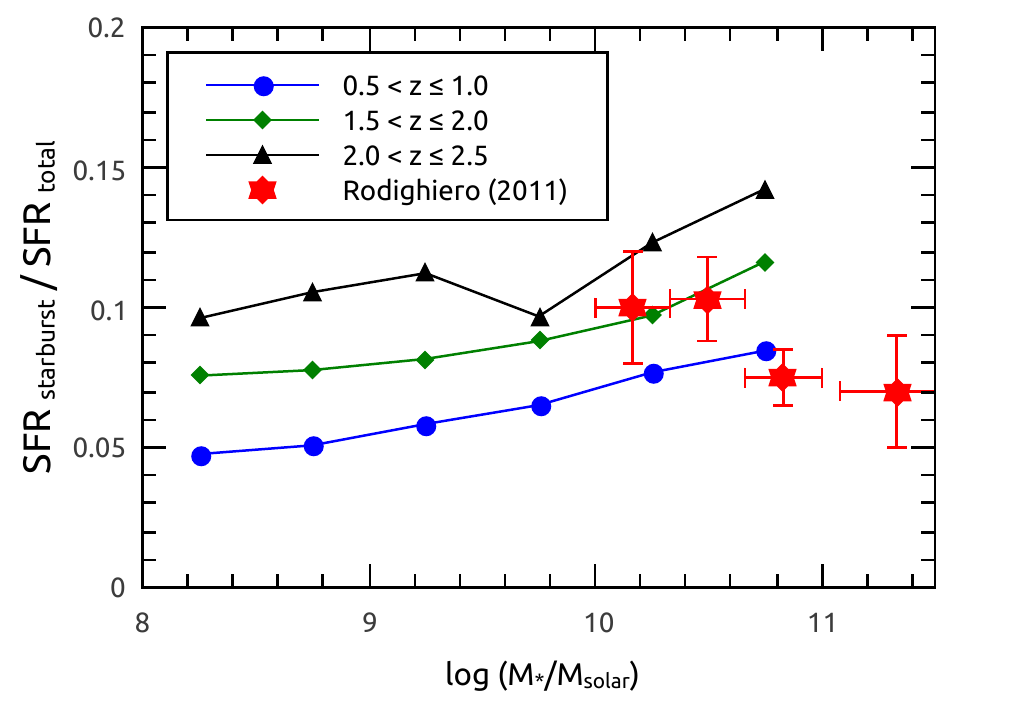}   
\caption{Contribution of starburst galaxies ($2.4\sigma$ above the MS) to the total SFR
as a function of stellar mass at different redshift bins. The observed \citet{2011ApJ...739L..40R}
data sets (red points with error bars) at $1.5<z<2.5$ are also plotted for comparison.}
\label{fig:sfrd}
\end{figure}

We quantify the merger-induced fraction of SFR by computing the
excess SFR owing to mergers from each histogram in Figure~\ref{fig:sfrd}.
To compare with data, we adopt a cut of $\sim0.45-0.5$ dex (or
$2.4\sigma$) above the MS (vertical dotted line), and declare the
fraction of SFR occurring above this to be merger-induced; this is
analogous to the procedure done in observations by
\cite{2011ApJ...739L..40R}.  The resulting values are very similar
to that obtained by directly summing the difference between the
histograms and the best-fit Gaussians in Figure~\ref{fig:PDF}.

Figure~\ref{fig:sfrd} shows that the amount of merger-induced SF
is generally quite small, around $\sim 5-15\%$, at all epochs out
to $z\sim 2.5$.  There is a mild trend for an increase in merger-induced
SF towards higher redshifts.  Since these galaxies are contributing
disproportionately to the SFR relative to their numbers, the fraction
of galaxies that lie in this merger regime is even smaller.

Our predicted values are quite consistent with the observational
determination by \cite{2011ApJ...739L..40R}, at the relevant redshift
$z\sim 2$.  Also we do not predict a strong mass dependence, which
is also generally consistent with these data.  Interestingly, we 
predict a weak increase in merger-induced SF to higher masses,
which is expected because halo mergers are increasingly important
at high halo masses~\citep{2008MNRAS.384....2G}.  The data if anything seem
to favor an anti-correlation with mass, but the dynamic range is
not yet large enough to make conclusive statements.
There are, however, some observational indications of bursty
star formation at lower masses on timescales much less than
$\sim 100$ Myr~\citep{2012ApJ...744...44W,2014MNRAS.441.2717K,
2016arXiv160405314G}. Even though these bursty population of
low-mass galaxies can contribute as much as $50-60\%$ to their
present-day mass~\citep{2014MNRAS.441.2717K}, the equilibrium
model does not predict any such trend.

Overall, during the peak activity of cosmic star formation until
today, galaxy growth in our equilibrium model is strongly dominated
by galaxies lying within the Gaussian scatter around the main
sequence.  This conclusion is consistent with currently available
observations, as well as longstanding predictions from cosmological
simulations~\citep{2002ApJ...571....1M,2005MNRAS.363....2K}.  This motivates
the idea that merger-induced star formation represent a second-order
effect in global galaxy growth, while the primary driver remains steady
(but mildly fluctuating) gravitational inflow.

\subsection{Timescale variability in SFRs}\label{subsec:timescale_variability}

So far, we have estimated the dispersion in MS using the SFRs
averaged over $100$ Myr.  However, the inferred SFRs of galaxies
are sensitive to the choice of timescale over which the SFR is
averaged.
\citep{2012ARA&A..50..531K,2014MNRAS.445..581H,2015MNRAS.447.3548S}.
H$\alpha$ tends to measure star formation traced by the most massive
stars hence and fluctuations on scales of tens of Myr, while UV-based
measures tend to trace somewhat less massive (OB) stars with
timescales of $\sim 100$~Myr.  Far-infrared measures come from
dust-reprocessed light which can trace even longer timescales.
Hence it is interesting to measure the scatter in SFR smoothed over
different time intervals.

\begin{figure}
  \includegraphics[width=0.5\textwidth,keepaspectratio]{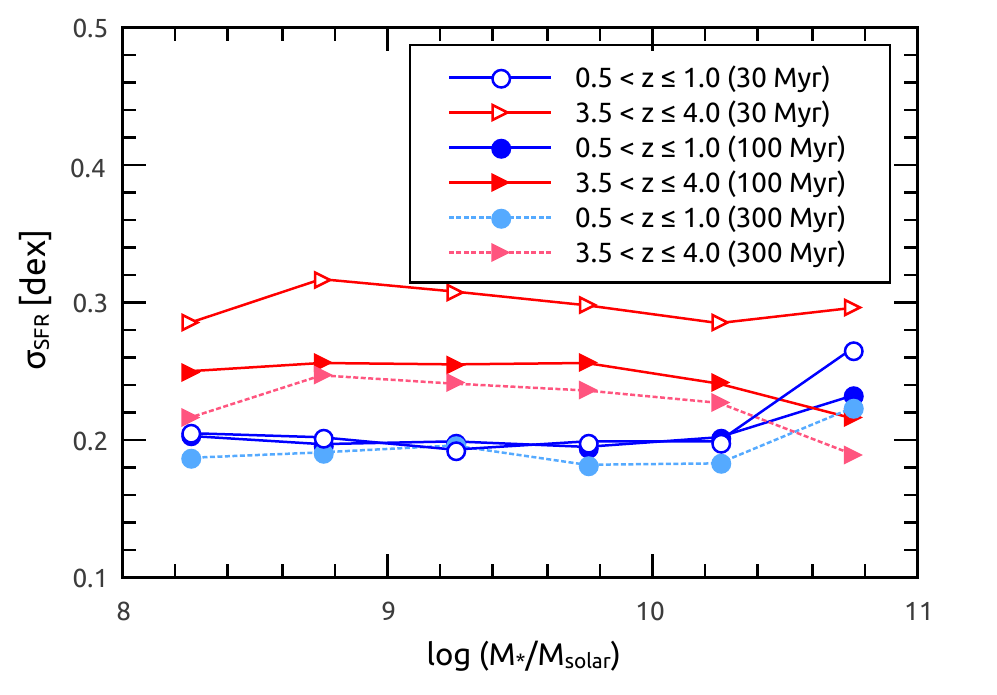}   
\caption{Scatter in main sequence with the SFR averaged over different timescales
($30$, $100$ and $300$ Myr). Overall, the dispersion increases for shorter
timescale variability and declines for longer timescales.}
\label{fig:sigma_dt}
\end{figure}

Figure~\ref{fig:sigma_dt} shows the SFR scatter averaged over $30$
Myr (also see Table~\ref{tab:sigma}) and $300$ Myr, along with our
canonical value of 100~Myr, for two different redshift ranges: $0.5<
z\leqslant1$ (circle) and $3.5< z\leqslant4$ (right triangle).
Generally, the scatter decreases when we increase the timescale,
which is expected as the SFHs of galaxies are essentially smoothed
out for longer timescale variability.  This is qualitatively in
agreement with other studies
\citep{2014MNRAS.445..581H,2015MNRAS.447.3548S}.

In detail, at low redshifts the differences among timescales are
fairly negligible.  This is expected because the accretion timescales
at late epochs are quite long.  Meanwhile, at high redshifts there
is a very clear trend, with 30~Myr scatter being $\approx 0.3$~dex
as opposed to longer timescale scatter being $\approx 0.25$~dex.
Still, all these variations are rather modest.  Hence the equilibrium
model does not predict that short timescale SFR indicators such as
H$\alpha$ will display significantly greater scatter.  It remains
to be seen if this is in agreement with observations.

\section{Summary}\label{sec:summary}

The equilibrium model for galaxy evolution highlights galactic
inflows and outflows as the main governor of galaxy growth, in
accord with the emerging baryon cycling paradigm for galaxy
self-regulation. Using a Bayesian Monte Carlo Markov chain approach,
we showed in Paper~I that our baryon cycling-based model can match
observed mean galaxy scaling relations across the majority of cosmic
time better and with many fewer parameters than in traditional
merger-tree based semi-analytic models. 

In this work, we test the basic equilibrium model by capturing the
``second-order'' galaxy evolutionary processes i.e. the deviations
from these mean trends, which we assume here are driven by the
fluctuations in the inflow rate, including mergers. The goal is to
understand the origin of the scatter in the scaling relation between
the stellar mass and the star formation rate (known as the main
sequence), and quantify the contribution from the scatter in mean
dark matter accretion rates.

We introduce a novel approach to incorporate fluctuations in the
inflow rates by sampling the merger rates derived from N-body
simulations.  Assuming that such dark matter fluctuations in halo
accretion are proportionally translated into baryonic inflow
fluctuations into the galaxy, we make predictions for the scatter
around the star forming galaxy main sequence.  We calculate the
1-$\sigma$ scatter within stellar mass bins spanning from
$8<\log(M_*/M_\odot)<11$ for different redshift ranges, by fitting
a Gaussian to the log of the SFR distribution (i.e. a log-normal).
Although the distributions of $\log{\rm SFR}$ are somewhat skewed
with an excess tail towards large SFR, they can still be reasonably
described by a single Gaussian fit \citep{2011ApJ...739L..40R} that
encloses the MS.  The tail portion corresponds to the starburst
regime.  We average over several different timescales to estimate
how the scatter would change amongst different SFR indicators.  In
this way, we quantify the intrinsic MS scatter as a function of
mass, redshift, and timescale.

Our main findings are summarized as follows:
\begin{itemize}

\item With a sufficiently large number of realizations, we show
that the linearly averaged mean of that inflow rate distribution
closely follows our the smooth accretion rate we assumed in Paper~I.
Hence we confirm that our stochastic approach closely reproduces
the results from our first-order model in Paper~I.  For higher
redshifts, they are almost identical as we enter a regime of smooth
accretion mode, while to lower redshifts the contribution from
mergers grows (though is still sub-dominant).

\item The merger-based inflow fluctuations translate into a significant
intrinsic scatter in the star-forming main sequence of $\approx
0.2-0.25$ dex.  This is generally somewhat lower than observed
values of the scatter which include measurement error, and is
comparable to the lowest values when observations attempt to infer
the intrinsic scatter by subtracting off measurement error.
Interestingly, our reported value is in excellent agreement with
the results from \cite{2014ApJS..214...15S} where they compile
various observed data sets and find the true scatter in MS, after
correcting for the observation-induced errors, to be $\sim 0.2$ dex rather
than $\sim 0.3$ dex.
This highlights our primary result, that fluctuations in the dark matter
accretion rate are the primary driver of the observed MS scatter.

\item We predict very little dependence on stellar mass in the
scatter down to $M_*=10^9 M_\odot$.  this trend is broadly in
agreement with observational studies
\citep{2012ApJ...754L..29W,2015A&A...575A..74S,2016MNRAS.455.2592R,2016ApJ...820L...1K}.
It is also generally consistent with hydrodynamic simulations, but
is less consistent with the higher-resolution FIRE simulations that
show must burstier star formation at low masses owing to a duty
cycle set up by intermittent strong feedback.  If such burstiness
is eventually confirmed in observations, it would suggest that we
must include variability in our mass outflow rate $\eta$ at low
masses into our equilibrium model. Currently, however, most
of the observations do not conclusively favour this (but also see
\citealt{2012ApJ...744...44W,2014MNRAS.441.2717K,2016arXiv160405314G}).

\item We predict that the MS scatter has only minimal redshift
dependence, increasing slightly at high redshifts.  We also show
that at later epochs, the scatter is mostly independent of SFR
timescale, while at early epochs it increases modestly to shorter
timescales.

\item The contribution to the global SFR from merger-induced star
formation is minimal at all explored redshifts, typically $5-15$\%
with a mild trend towards being higher at high redshifts.  Hence
starbursts are strongly sub-dominant in terms of overall stellar
growth in our models.  This is consistent with available observations
and cosmologically-situated models.

\end{itemize}

Our results indicate that we have successfully tested the equilibrium
model by predicting scatter around the main sequence purely from inflow
fluctuations, and that such fluctuations are the primary driver
for deviations from the main sequence.

In this work we have not considered fluctuations in the baryon
cycling parameters.  Potentially, one could use the scatter in the
mass-metallicity relation to constrain this.  In the absence of
recycling, the mass dependence of the metallicity is purely set by
that of $\eta$, so in principle the mass-metallicity scatter
constrains the $\eta$ scatter.  But one could likely also introduce
scatter in $\zeta$ and/or $t_{\rm rec}$ to achieve a similar result.
Satellites also deviate systematically in metallicity from central
galaxies, so we are engaged in developing models to incorporate
satellite stripping processes in order to quantify their contribution
to the MS scatter (Simha et al., in preparation).  Still, given
that the scatter in mass-metallicity is typically quite small ($\sim
0.1$~dex), this suggests that inflow fluctuations will continue to
dominate the MS scatter.

The equilibrium model is developing into a valuable and intuitive
tool to study galaxy evolution within a simple baryon cycling
framework.  In a sense, this model is starting down the road to
``precision galaxy formation", in which the first-order parameters
that establish galaxy growth can start to be constrained by
observations, in analogy with the main parameters in precision
cosmology.  Meanwhile, other processes such as galaxy mergers
represent second-order effects to global galaxy growth, and impact
a different set of observables such as galaxy morphological
transformations and satellite-specific processes.  Incorporating
halo mergers in a probabilistic fashion as we have done here is a
key step towards extensions of the equilibrium model that will
develop this tool into a leading platform for understanding galaxy
evolution within a baryon cycling framework.

\section*{Acknowledgements}
SM, RD, and VS acknowledge support from the South African Research
Chairs Initiative and the South African National Research Foundation.
Support for RD was also provided by NASA ATP grant NNX12AH86G to the
University of Arizona. 

\bibliography{mitra}

\begin{thebibliography}{}
\makeatletter
\relax
\def\mn@urlcharsother{\let\do\@makeother \do\$\do\&\do\#\do\^\do\_\do\%\do\~}
\def\mn@doi{\begingroup\mn@urlcharsother \@ifnextchar [ {\mn@doi@}
  {\mn@doi@[]}}
\def\mn@doi@[#1]#2{\def\@tempa{#1}\ifx\@tempa\@empty \href
  {http://dx.doi.org/#2} {doi:#2}\else \href {http://dx.doi.org/#2} {#1}\fi
  \endgroup}
\def\mn@eprint#1#2{\mn@eprint@#1:#2::\@nil}
\def\mn@eprint@arXiv#1{\href {http://arxiv.org/abs/#1} {{\tt arXiv:#1}}}
\def\mn@eprint@dblp#1{\href {http://dblp.uni-trier.de/rec/bibtex/#1.xml}
  {dblp:#1}}
\def\mn@eprint@#1:#2:#3:#4\@nil{\def\@tempa {#1}\def\@tempb {#2}\def\@tempc
  {#3}\ifx \@tempc \@empty \let \@tempc \@tempb \let \@tempb \@tempa \fi \ifx
  \@tempb \@empty \def\@tempb {arXiv}\fi \@ifundefined
  {mn@eprint@\@tempb}{\@tempb:\@tempc}{\expandafter \expandafter \csname
  mn@eprint@\@tempb\endcsname \expandafter{\@tempc}}}

\bibitem[\protect\citeauthoryear{{Andrews} \& {Martini}}{{Andrews} \&
  {Martini}}{2013}]{2013ApJ...765..140A}
{Andrews} B.~H.,  {Martini} P.,  2013, \mn@doi [\apj]
  {10.1088/0004-637X/765/2/140}, \href
  {http://adsabs.harvard.edu/abs/2013ApJ...765..140A} {765, 140}

\bibitem[\protect\citeauthoryear{{Balogh}, {Pearce}, {Bower}  \&
  {Kay}}{{Balogh} et~al.}{2001}]{2001MNRAS.326.1228B}
{Balogh} M.~L.,  {Pearce} F.~R.,  {Bower} R.~G.,   {Kay} S.~T.,  2001, \mn@doi
  [\mnras] {10.1111/j.1365-2966.2001.04667.x}, \href
  {http://adsabs.harvard.edu/abs/2001MNRAS.326.1228B} {326, 1228}

\bibitem[\protect\citeauthoryear{{Behroozi}, {Wechsler}  \&
  {Conroy}}{{Behroozi} et~al.}{2013}]{2013ApJ...770...57B}
{Behroozi} P.~S.,  {Wechsler} R.~H.,   {Conroy} C.,  2013, \mn@doi [\apj]
  {10.1088/0004-637X/770/1/57}, \href
  {http://adsabs.harvard.edu/abs/2013ApJ...770...57B} {770, 57}

\bibitem[\protect\citeauthoryear{{Bouch{\'e}} et~al.,}{{Bouch{\'e}}
  et~al.}{2010}]{2010ApJ...718.1001B}
{Bouch{\'e}} N.,  et~al., 2010, \mn@doi [\apj] {10.1088/0004-637X/718/2/1001},
  \href {http://adsabs.harvard.edu/abs/2010ApJ...718.1001B} {718, 1001}

\bibitem[\protect\citeauthoryear{{Boylan-Kolchin}, {Springel}, {White},
  {Jenkins}  \& {Lemson}}{{Boylan-Kolchin} et~al.}{2009}]{2009MNRAS.398.1150B}
{Boylan-Kolchin} M.,  {Springel} V.,  {White} S.~D.~M.,  {Jenkins} A.,
  {Lemson} G.,  2009, \mn@doi [\mnras] {10.1111/j.1365-2966.2009.15191.x},
  \href {http://adsabs.harvard.edu/abs/2009MNRAS.398.1150B} {398, 1150}

\bibitem[\protect\citeauthoryear{{Christensen}, {Dav{\'e}}, {Governato},
  {Pontzen}, {Brooks}, {Munshi}, {Quinn}  \& {Wadsley}}{{Christensen}
  et~al.}{2016}]{2016ApJ...824...57C}
{Christensen} C.~R.,  {Dav{\'e}} R.,  {Governato} F.,  {Pontzen} A.,  {Brooks}
  A.,  {Munshi} F.,  {Quinn} T.,   {Wadsley} J.,  2016, \mn@doi [\apj]
  {10.3847/0004-637X/824/1/57}, \href
  {http://adsabs.harvard.edu/abs/2016ApJ...824...57C} {824, 57}

\bibitem[\protect\citeauthoryear{{Cox}, {Jonsson}, {Somerville}, {Primack}  \&
  {Dekel}}{{Cox} et~al.}{2008}]{2008MNRAS.384..386C}
{Cox} T.~J.,  {Jonsson} P.,  {Somerville} R.~S.,  {Primack} J.~R.,   {Dekel}
  A.,  2008, \mn@doi [\mnras] {10.1111/j.1365-2966.2007.12730.x}, \href
  {http://adsabs.harvard.edu/abs/2008MNRAS.384..386C} {384, 386}

\bibitem[\protect\citeauthoryear{{Daddi} et~al.}{{Daddi}
  et~al.}{2007}]{2007ApJ...670..156D}
{Daddi} E.,  et~al., 2007, \mn@doi [\apj] {10.1086/521818}, \href
  {http://adsabs.harvard.edu/abs/2007ApJ...670..156D} {670, 156}

\bibitem[\protect\citeauthoryear{{Dav{\'e}}, {Oppenheimer}  \&
  {Finlator}}{{Dav{\'e}} et~al.}{2011}]{dav11a}
{Dav{\'e}} R.,  {Oppenheimer} B.~D.,   {Finlator} K.,  2011, \mn@doi [\mnras]
  {10.1111/j.1365-2966.2011.18680.x}, \href
  {http://adsabs.harvard.edu/abs/2011MNRAS.415...11D} {415, 11}

\bibitem[\protect\citeauthoryear{{Dav{\'e}}, {Finlator}  \&
  {Oppenheimer}}{{Dav{\'e}} et~al.}{2012}]{dav12}
{Dav{\'e}} R.,  {Finlator} K.,   {Oppenheimer} B.~D.,  2012, \mn@doi [\mnras]
  {10.1111/j.1365-2966.2011.20148.x}, \href
  {http://adsabs.harvard.edu/abs/2012MNRAS.421...98D} {421, 98}

\bibitem[\protect\citeauthoryear{{Dav{\'e}}, {Thompson}  \&
  {Hopkins}}{{Dav{\'e}} et~al.}{2016}]{dav16}
{Dav{\'e}} R.,  {Thompson} R.,   {Hopkins} P.~F.,  2016, \mn@doi [\mnras]
  {10.1093/mnras/stw1862}, \href
  {http://adsabs.harvard.edu/abs/2016MNRAS.462.3265D} {462, 3265}

\bibitem[\protect\citeauthoryear{{Dekel} \& {Mandelker}}{{Dekel} \&
  {Mandelker}}{2014}]{2014MNRAS.444.2071D}
{Dekel} A.,  {Mandelker} N.,  2014, \mn@doi [\mnras] {10.1093/mnras/stu1427},
  \href {http://adsabs.harvard.edu/abs/2014MNRAS.444.2071D} {444, 2071}

\bibitem[\protect\citeauthoryear{{Dekel} et~al.}{{Dekel}
  et~al.}{2009}]{2009Natur.457..451D}
{Dekel} A.,  et~al., 2009, \mn@doi [\nat] {10.1038/nature07648}, \href
  {http://adsabs.harvard.edu/abs/2009Natur.457..451D} {457, 451}

\bibitem[\protect\citeauthoryear{{Di Cintio}, {Brook}, {Macci{\`o}}, {Stinson},
  {Knebe}, {Dutton}  \& {Wadsley}}{{Di Cintio}
  et~al.}{2014}]{2014MNRAS.437..415D}
{Di Cintio} A.,  {Brook} C.~B.,  {Macci{\`o}} A.~V.,  {Stinson} G.~S.,  {Knebe}
  A.,  {Dutton} A.~A.,   {Wadsley} J.,  2014, \mn@doi [\mnras]
  {10.1093/mnras/stt1891}, \href
  {http://adsabs.harvard.edu/abs/2014MNRAS.437..415D} {437, 415}

\bibitem[\protect\citeauthoryear{{Dutton}, {van den Bosch}  \&
  {Dekel}}{{Dutton} et~al.}{2010}]{2010MNRAS.405.1690D}
{Dutton} A.~A.,  {van den Bosch} F.~C.,   {Dekel} A.,  2010, \mn@doi [\mnras]
  {10.1111/j.1365-2966.2010.16620.x}, \href
  {http://adsabs.harvard.edu/abs/2010MNRAS.405.1690D} {405, 1690}

\bibitem[\protect\citeauthoryear{{Elbaz} et~al.,}{{Elbaz}
  et~al.}{2007}]{2007A&A...468...33E}
{Elbaz} D.,  et~al., 2007, \mn@doi [\aap] {10.1051/0004-6361:20077525}, \href
  {http://adsabs.harvard.edu/abs/2007A%26A...468...33E} {468, 33}

\bibitem[\protect\citeauthoryear{{Fakhouri}, {Ma}  \&
  {Boylan-Kolchin}}{{Fakhouri} et~al.}{2010}]{2010MNRAS.406.2267F}
{Fakhouri} O.,  {Ma} C.-P.,   {Boylan-Kolchin} M.,  2010, \mn@doi [\mnras]
  {10.1111/j.1365-2966.2010.16859.x}, \href
  {http://adsabs.harvard.edu/abs/2010MNRAS.406.2267F} {406, 2267}

\bibitem[\protect\citeauthoryear{{Finlator} \& {Dav{\'e}}}{{Finlator} \&
  {Dav{\'e}}}{2008}]{2008MNRAS.385.2181F}
{Finlator} K.,  {Dav{\'e}} R.,  2008, \mn@doi [\mnras]
  {10.1111/j.1365-2966.2008.12991.x}, \href
  {http://adsabs.harvard.edu/abs/2008MNRAS.385.2181F} {385, 2181}

\bibitem[\protect\citeauthoryear{{Forbes}, {Krumholz}, {Burkert}  \&
  {Dekel}}{{Forbes} et~al.}{2014}]{2014MNRAS.443..168F}
{Forbes} J.~C.,  {Krumholz} M.~R.,  {Burkert} A.,   {Dekel} A.,  2014, \mn@doi
  [\mnras] {10.1093/mnras/stu1142}, \href
  {http://adsabs.harvard.edu/abs/2014MNRAS.443..168F} {443, 168}

\bibitem[\protect\citeauthoryear{{Gabor} \& {Dav{\'e}}}{{Gabor} \&
  {Dav{\'e}}}{2015}]{2015MNRAS.447..374G}
{Gabor} J.~M.,  {Dav{\'e}} R.,  2015, \mn@doi [\mnras] {10.1093/mnras/stu2399},
  \href {http://adsabs.harvard.edu/abs/2015MNRAS.447..374G} {447, 374}

\bibitem[\protect\citeauthoryear{{Genel} et~al.,}{{Genel}
  et~al.}{2014}]{2014MNRAS.445..175G}
{Genel} S.,  et~al., 2014, \mn@doi [\mnras] {10.1093/mnras/stu1654}, \href
  {http://adsabs.harvard.edu/abs/2014MNRAS.445..175G} {445, 175}

\bibitem[\protect\citeauthoryear{{Goerdt}, {Ceverino}, {Dekel}  \&
  {Teyssier}}{{Goerdt} et~al.}{2015}]{2015MNRAS.454..637G}
{Goerdt} T.,  {Ceverino} D.,  {Dekel} A.,   {Teyssier} R.,  2015, \mn@doi
  [\mnras] {10.1093/mnras/stv2005}, \href
  {http://adsabs.harvard.edu/abs/2015MNRAS.454..637G} {454, 637}

\bibitem[\protect\citeauthoryear{{Guo} \& {White}}{{Guo} \&
  {White}}{2008}]{2008MNRAS.384....2G}
{Guo} Q.,  {White} S.~D.~M.,  2008, \mn@doi [\mnras]
  {10.1111/j.1365-2966.2007.12619.x}, \href
  {http://adsabs.harvard.edu/abs/2008MNRAS.384....2G} {384, 2}

\bibitem[\protect\citeauthoryear{{Guo}, {Zheng}  \& {Fu}}{{Guo}
  et~al.}{2013}]{2013ApJ...778...23G}
{Guo} K.,  {Zheng} X.~Z.,   {Fu} H.,  2013, \mn@doi [\apj]
  {10.1088/0004-637X/778/1/23}, \href
  {http://adsabs.harvard.edu/abs/2013ApJ...778...23G} {778, 23}

\bibitem[\protect\citeauthoryear{{Guo} et~al.,}{{Guo}
  et~al.}{2016}]{2016arXiv160405314G}
{Guo} Y.,  et~al., 2016, preprint, \href
  {http://adsabs.harvard.edu/abs/2016arXiv160405314G} {} (\mn@eprint {arXiv}
  {1604.05314})

\bibitem[\protect\citeauthoryear{{Henriques}, {White}, {Thomas}, {Angulo},
  {Guo}, {Lemson}  \& {Springel}}{{Henriques}
  et~al.}{2013}]{2013MNRAS.431.3373H}
{Henriques} B.~M.~B.,  {White} S.~D.~M.,  {Thomas} P.~A.,  {Angulo} R.~E.,
  {Guo} Q.,  {Lemson} G.,   {Springel} V.,  2013, \mn@doi [\mnras]
  {10.1093/mnras/stt415}, \href
  {http://adsabs.harvard.edu/abs/2013MNRAS.431.3373H} {431, 3373}

\bibitem[\protect\citeauthoryear{{Hopkins}, {Kere{\v s}}, {O{\~n}orbe},
  {Faucher-Gigu{\`e}re}, {Quataert}, {Murray}  \& {Bullock}}{{Hopkins}
  et~al.}{2014}]{2014MNRAS.445..581H}
{Hopkins} P.~F.,  {Kere{\v s}} D.,  {O{\~n}orbe} J.,  {Faucher-Gigu{\`e}re}
  C.-A.,  {Quataert} E.,  {Murray} N.,   {Bullock} J.~S.,  2014, \mn@doi
  [\mnras] {10.1093/mnras/stu1738}, \href
  {http://adsabs.harvard.edu/abs/2014MNRAS.445..581H} {445, 581}

\bibitem[\protect\citeauthoryear{{Johnston}, {Vaccari}, {Jarvis}, {Smith},
  {Giovannoli}, {H{\"a}u{\ss}ler}  \& {Prescott}}{{Johnston}
  et~al.}{2015}]{2015MNRAS.453.2540J}
{Johnston} R.,  {Vaccari} M.,  {Jarvis} M.,  {Smith} M.,  {Giovannoli} E.,
  {H{\"a}u{\ss}ler} B.,   {Prescott} M.,  2015, \mn@doi [\mnras]
  {10.1093/mnras/stv1715}, \href
  {http://adsabs.harvard.edu/abs/2015MNRAS.453.2540J} {453, 2540}

\bibitem[\protect\citeauthoryear{{Kauffmann}}{{Kauffmann}}{2014}]{2014MNRAS.441.2717K}
{Kauffmann} G.,  2014, \mn@doi [\mnras] {10.1093/mnras/stu752}, \href
  {http://adsabs.harvard.edu/abs/2014MNRAS.441.2717K} {441, 2717}

\bibitem[\protect\citeauthoryear{{Kennicutt} \& {Evans}}{{Kennicutt} \&
  {Evans}}{2012}]{2012ARA&A..50..531K}
{Kennicutt} R.~C.,  {Evans} N.~J.,  2012, \mn@doi [\araa]
  {10.1146/annurev-astro-081811-125610}, \href
  {http://adsabs.harvard.edu/abs/2012ARA%26A..50..531K} {50, 531}

\bibitem[\protect\citeauthoryear{{Kere{\v s}}, {Katz}, {Weinberg}  \&
  {Dav{\'e}}}{{Kere{\v s}} et~al.}{2005}]{2005MNRAS.363....2K}
{Kere{\v s}} D.,  {Katz} N.,  {Weinberg} D.~H.,   {Dav{\'e}} R.,  2005, \mn@doi
  [\mnras] {10.1111/j.1365-2966.2005.09451.x}, \href
  {http://adsabs.harvard.edu/abs/2005MNRAS.363....2K} {363, 2}

\bibitem[\protect\citeauthoryear{{Krumholz} \& {Dekel}}{{Krumholz} \&
  {Dekel}}{2012}]{2012ApJ...753...16K}
{Krumholz} M.~R.,  {Dekel} A.,  2012, \mn@doi [\apj]
  {10.1088/0004-637X/753/1/16}, \href
  {http://adsabs.harvard.edu/abs/2012ApJ...753...16K} {753, 16}

\bibitem[\protect\citeauthoryear{{Kurczynski} et~al.,}{{Kurczynski}
  et~al.}{2016}]{2016ApJ...820L...1K}
{Kurczynski} P.,  et~al., 2016, \mn@doi [\apjl] {10.3847/2041-8205/820/1/L1},
  \href {http://adsabs.harvard.edu/abs/2016ApJ...820L...1K} {820, L1}

\bibitem[\protect\citeauthoryear{{Lilly}, {Carollo}, {Pipino}, {Renzini}  \&
  {Peng}}{{Lilly} et~al.}{2013}]{2013ApJ...772..119L}
{Lilly} S.~J.,  {Carollo} C.~M.,  {Pipino} A.,  {Renzini} A.,   {Peng} Y.,
  2013, \mn@doi [\apj] {10.1088/0004-637X/772/2/119}, \href
  {http://adsabs.harvard.edu/abs/2013ApJ...772..119L} {772, 119}

\bibitem[\protect\citeauthoryear{{Martin}}{{Martin}}{2005}]{2005ApJ...621..227M}
{Martin} C.~L.,  2005, \mn@doi [\apj] {10.1086/427277}, \href
  {http://adsabs.harvard.edu/abs/2005ApJ...621..227M} {621, 227}

\bibitem[\protect\citeauthoryear{{Mihos} \& {Hernquist}}{{Mihos} \&
  {Hernquist}}{1996}]{1996ApJ...464..641M}
{Mihos} J.~C.,  {Hernquist} L.,  1996, \mn@doi [\apj] {10.1086/177353}, \href
  {http://adsabs.harvard.edu/abs/1996ApJ...464..641M} {464, 641}

\bibitem[\protect\citeauthoryear{{Mitra}, {Dav{\'e}}  \& {Finlator}}{{Mitra}
  et~al.}{2015}]{Mitra15}
{Mitra} S.,  {Dav{\'e}} R.,   {Finlator} K.,  2015, \mn@doi [\mnras]
  {10.1093/mnras/stv1387}, \href
  {http://adsabs.harvard.edu/abs/2015MNRAS.452.1184M} {452, 1184}

\bibitem[\protect\citeauthoryear{{Mo}, {Mao}  \& {White}}{{Mo}
  et~al.}{1998}]{1998MNRAS.295..319M}
{Mo} H.~J.,  {Mao} S.,   {White} S.~D.~M.,  1998, \mn@doi [\mnras]
  {10.1046/j.1365-8711.1998.01227.x}, \href
  {http://adsabs.harvard.edu/abs/1998MNRAS.295..319M} {295, 319}

\bibitem[\protect\citeauthoryear{{Moster}, {Naab}  \& {White}}{{Moster}
  et~al.}{2013}]{2013MNRAS.428.3121M}
{Moster} B.~P.,  {Naab} T.,   {White} S.~D.~M.,  2013, \mn@doi [\mnras]
  {10.1093/mnras/sts261}, \href
  {http://adsabs.harvard.edu/abs/2013MNRAS.428.3121M} {428, 3121}

\bibitem[\protect\citeauthoryear{{Murali}, {Katz}, {Hernquist}, {Weinberg}  \&
  {Dav{\'e}}}{{Murali} et~al.}{2002}]{2002ApJ...571....1M}
{Murali} C.,  {Katz} N.,  {Hernquist} L.,  {Weinberg} D.~H.,   {Dav{\'e}} R.,
  2002, \mn@doi [\apj] {10.1086/339876}, \href
  {http://adsabs.harvard.edu/abs/2002ApJ...571....1M} {571, 1}

\bibitem[\protect\citeauthoryear{{Muratov}, {Kere{\v s}},
  {Faucher-Gigu{\`e}re}, {Hopkins}, {Quataert}  \& {Murray}}{{Muratov}
  et~al.}{2015}]{2015MNRAS.454.2691M}
{Muratov} A.~L.,  {Kere{\v s}} D.,  {Faucher-Gigu{\`e}re} C.-A.,  {Hopkins}
  P.~F.,  {Quataert} E.,   {Murray} N.,  2015, \mn@doi [\mnras]
  {10.1093/mnras/stv2126}, \href
  {http://adsabs.harvard.edu/abs/2015MNRAS.454.2691M} {454, 2691}

\bibitem[\protect\citeauthoryear{{Noeske} et~al.}{{Noeske}
  et~al.}{2007}]{2007ApJ...660L..43N}
{Noeske} K.~G.,  et~al., 2007, \mn@doi [\apjl.] {10.1086/517926}, \href
  {http://adsabs.harvard.edu/abs/2007ApJ...660L..43N} {660, L43}

\bibitem[\protect\citeauthoryear{{Oppenheimer}, {Dav{\'e}}, {Kere{\v s}},
  {Fardal}, {Katz}, {Kollmeier}  \& {Weinberg}}{{Oppenheimer}
  et~al.}{2010}]{2010MNRAS.406.2325O}
{Oppenheimer} B.~D.,  {Dav{\'e}} R.,  {Kere{\v s}} D.,  {Fardal} M.,  {Katz}
  N.,  {Kollmeier} J.~A.,   {Weinberg} D.~H.,  2010, \mn@doi [\mnras]
  {10.1111/j.1365-2966.2010.16872.x}, \href
  {http://adsabs.harvard.edu/abs/2010MNRAS.406.2325O} {406, 2325}

\bibitem[\protect\citeauthoryear{{Peng} \& {Maiolino}}{{Peng} \&
  {Maiolino}}{2014}]{2014MNRAS.443.3643P}
{Peng} Y.-j.,  {Maiolino} R.,  2014, \mn@doi [\mnras] {10.1093/mnras/stu1288},
  \href {http://adsabs.harvard.edu/abs/2014MNRAS.443.3643P} {443, 3643}

\bibitem[\protect\citeauthoryear{{Planck Collaboration} et~al.,}{{Planck
  Collaboration} et~al.}{2015}]{2015arXiv150201589P}
{Planck Collaboration} et~al., 2015, preprint, \href
  {http://adsabs.harvard.edu/abs/2015arXiv150201589P} {} (\mn@eprint {arXiv}
  {1502.01589})

\bibitem[\protect\citeauthoryear{{Rodighiero} et~al.}{{Rodighiero}
  et~al.}{2011}]{2011ApJ...739L..40R}
{Rodighiero} G.,  et~al., 2011, \mn@doi [\apjl.] {10.1088/2041-8205/739/2/L40},
  \href {http://adsabs.harvard.edu/abs/2011ApJ...739L..40R} {739, L40}

\bibitem[\protect\citeauthoryear{{Rodr{\'{\i}}guez-Puebla}, {Primack},
  {Behroozi}  \& {Faber}}{{Rodr{\'{\i}}guez-Puebla}
  et~al.}{2016}]{2016MNRAS.455.2592R}
{Rodr{\'{\i}}guez-Puebla} A.,  {Primack} J.~R.,  {Behroozi} P.,   {Faber}
  S.~M.,  2016, \mn@doi [\mnras] {10.1093/mnras/stv2513}, \href
  {http://adsabs.harvard.edu/abs/2016MNRAS.455.2592R} {455, 2592}

\bibitem[\protect\citeauthoryear{{Rubin}, {Prochaska}, {Koo}, {Phillips},
  {Martin}  \& {Winstrom}}{{Rubin} et~al.}{2014}]{2014ApJ...794..156R}
{Rubin} K.~H.~R.,  {Prochaska} J.~X.,  {Koo} D.~C.,  {Phillips} A.~C.,
  {Martin} C.~L.,   {Winstrom} L.~O.,  2014, \mn@doi [\apj]
  {10.1088/0004-637X/794/2/156}, \href
  {http://adsabs.harvard.edu/abs/2014ApJ...794..156R} {794, 156}

\bibitem[\protect\citeauthoryear{{Saintonge} et~al.}{{Saintonge}
  et~al.}{2013}]{2013ApJ...778....2S}
{Saintonge} A.,  et~al., 2013, \mn@doi [\apj] {10.1088/0004-637X/778/1/2},
  \href {http://adsabs.harvard.edu/abs/2013ApJ...778....2S} {778, 2}

\bibitem[\protect\citeauthoryear{{Salim} et~al.}{{Salim}
  et~al.}{2007}]{2007ApJS..173..267S}
{Salim} S.,  et~al., 2007, \mn@doi [\apjs] {10.1086/519218}, \href
  {http://adsabs.harvard.edu/abs/2007ApJS..173..267S} {173, 267}

\bibitem[\protect\citeauthoryear{{Sanders} \& {Mirabel}}{{Sanders} \&
  {Mirabel}}{1996}]{1996ARA&A..34..749S}
{Sanders} D.~B.,  {Mirabel} I.~F.,  1996, \mn@doi [\araa]
  {10.1146/annurev.astro.34.1.749}, \href
  {http://adsabs.harvard.edu/abs/1996ARA%26A..34..749S} {34, 749}

\bibitem[\protect\citeauthoryear{{Sanders} et~al.,}{{Sanders}
  et~al.}{2015}]{2015ApJ...799..138S}
{Sanders} R.~L.,  et~al., 2015, \mn@doi [\apj] {10.1088/0004-637X/799/2/138},
  \href {http://adsabs.harvard.edu/abs/2015ApJ...799..138S} {799, 138}

\bibitem[\protect\citeauthoryear{{Sargent}, {B{\'e}thermin}, {Daddi}  \&
  {Elbaz}}{{Sargent} et~al.}{2012}]{2012ApJ...747L..31S}
{Sargent} M.~T.,  {B{\'e}thermin} M.,  {Daddi} E.,   {Elbaz} D.,  2012, \mn@doi
  [\apjl] {10.1088/2041-8205/747/2/L31}, \href
  {http://adsabs.harvard.edu/abs/2012ApJ...747L..31S} {747, L31}

\bibitem[\protect\citeauthoryear{{Schaye} et~al.,}{{Schaye}
  et~al.}{2015}]{2015MNRAS.446..521S}
{Schaye} J.,  et~al., 2015, \mn@doi [\mnras] {10.1093/mnras/stu2058}, \href
  {http://adsabs.harvard.edu/abs/2015MNRAS.446..521S} {446, 521}

\bibitem[\protect\citeauthoryear{{Schreiber} et~al.,}{{Schreiber}
  et~al.}{2015}]{2015A&A...575A..74S}
{Schreiber} C.,  et~al., 2015, \mn@doi [\aap] {10.1051/0004-6361/201425017},
  \href {http://adsabs.harvard.edu/abs/2015A%26A...575A..74S} {575, A74}

\bibitem[\protect\citeauthoryear{{Shivaei} et~al.,}{{Shivaei}
  et~al.}{2015}]{2015ApJ...815...98S}
{Shivaei} I.,  et~al., 2015, \mn@doi [\apj] {10.1088/0004-637X/815/2/98}, \href
  {http://adsabs.harvard.edu/abs/2015ApJ...815...98S} {815, 98}

\bibitem[\protect\citeauthoryear{{Somerville} \& {Dav{\'e}}}{{Somerville} \&
  {Dav{\'e}}}{2015}]{2015ARA&A..53...51S}
{Somerville} R.~S.,  {Dav{\'e}} R.,  2015, \mn@doi [\araa]
  {10.1146/annurev-astro-082812-140951}, \href
  {http://adsabs.harvard.edu/abs/2015ARA%26A..53...51S} {53, 51}

\bibitem[\protect\citeauthoryear{{Sparre}, {Hayward}, {Feldmann},
  {Faucher-Gigu{\`e}re}, {Muratov}, {Kere{\v s}}  \& {Hopkins}}{{Sparre}
  et~al.}{2015a}]{2015arXiv151003869S}
{Sparre} M.,  {Hayward} C.~C.,  {Feldmann} R.,  {Faucher-Gigu{\`e}re} C.-A.,
  {Muratov} A.~L.,  {Kere{\v s}} D.,   {Hopkins} P.~F.,  2015a, preprint, \href
  {http://adsabs.harvard.edu/abs/2015arXiv151003869S} {} (\mn@eprint {arXiv}
  {1510.03869})

\bibitem[\protect\citeauthoryear{{Sparre} et~al.,}{{Sparre}
  et~al.}{2015b}]{2015MNRAS.447.3548S}
{Sparre} M.,  et~al., 2015b, \mn@doi [\mnras] {10.1093/mnras/stu2713}, \href
  {http://adsabs.harvard.edu/abs/2015MNRAS.447.3548S} {447, 3548}

\bibitem[\protect\citeauthoryear{{Speagle}, {Steinhardt}, {Capak}  \&
  {Silverman}}{{Speagle} et~al.}{2014}]{2014ApJS..214...15S}
{Speagle} J.~S.,  {Steinhardt} C.~L.,  {Capak} P.~L.,   {Silverman} J.~D.,
  2014, \mn@doi [\apjs] {10.1088/0067-0049/214/2/15}, \href
  {http://adsabs.harvard.edu/abs/2014ApJS..214...15S} {214, 15}

\bibitem[\protect\citeauthoryear{{Springel} et~al.,}{{Springel}
  et~al.}{2005}]{2005Natur.435..629S}
{Springel} V.,  et~al., 2005, \mn@doi [\nat] {10.1038/nature03597}, \href
  {http://adsabs.harvard.edu/abs/2005Natur.435..629S} {435, 629}

\bibitem[\protect\citeauthoryear{{Steidel}, {Erb}, {Shapley}, {Pettini},
  {Reddy}, {Bogosavljevi{\'c}}, {Rudie}  \& {Rakic}}{{Steidel}
  et~al.}{2010}]{2010ApJ...717..289S}
{Steidel} C.~C.,  {Erb} D.~K.,  {Shapley} A.~E.,  {Pettini} M.,  {Reddy} N.,
  {Bogosavljevi{\'c}} M.,  {Rudie} G.~C.,   {Rakic} O.,  2010, \mn@doi [\apj]
  {10.1088/0004-637X/717/1/289}, \href
  {http://adsabs.harvard.edu/abs/2010ApJ...717..289S} {717, 289}

\bibitem[\protect\citeauthoryear{{Steidel} et~al.,}{{Steidel}
  et~al.}{2014}]{2014ApJ...795..165S}
{Steidel} C.~C.,  et~al., 2014, \mn@doi [\apj] {10.1088/0004-637X/795/2/165},
  \href {http://adsabs.harvard.edu/abs/2014ApJ...795..165S} {795, 165}

\bibitem[\protect\citeauthoryear{{Tacchella}, {Dekel}, {Carollo}, {Ceverino},
  {DeGraf}, {Lapiner}, {Mandelker}  \& {Primack Joel}}{{Tacchella}
  et~al.}{2016}]{2016MNRAS.457.2790T}
{Tacchella} S.,  {Dekel} A.,  {Carollo} C.~M.,  {Ceverino} D.,  {DeGraf} C.,
  {Lapiner} S.,  {Mandelker} N.,   {Primack Joel} R.,  2016, \mn@doi [\mnras]
  {10.1093/mnras/stw131}, \href
  {http://adsabs.harvard.edu/abs/2016MNRAS.457.2790T} {457, 2790}

\bibitem[\protect\citeauthoryear{{Tacconi} et~al.}{{Tacconi}
  et~al.}{2013}]{2013ApJ...768...74T}
{Tacconi} L.~J.,  et~al., 2013, \mn@doi [\apj] {10.1088/0004-637X/768/1/74},
  \href {http://adsabs.harvard.edu/abs/2013ApJ...768...74T} {768, 74}

\bibitem[\protect\citeauthoryear{{Tremonti} et~al.}{{Tremonti}
  et~al.}{2004}]{2004ApJ...613..898T}
{Tremonti} C.~A.,  et~al., 2004, \mn@doi [\apj] {10.1086/423264}, \href
  {http://adsabs.harvard.edu/abs/2004ApJ...613..898T} {613, 898}

\bibitem[\protect\citeauthoryear{{Weiner} et~al.,}{{Weiner}
  et~al.}{2009}]{2009ApJ...692..187W}
{Weiner} B.~J.,  et~al., 2009, \mn@doi [\apj] {10.1088/0004-637X/692/1/187},
  \href {http://adsabs.harvard.edu/abs/2009ApJ...692..187W} {692, 187}

\bibitem[\protect\citeauthoryear{{Weisz} et~al.,}{{Weisz}
  et~al.}{2012}]{2012ApJ...744...44W}
{Weisz} D.~R.,  et~al., 2012, \mn@doi [\apj] {10.1088/0004-637X/744/1/44},
  \href {http://adsabs.harvard.edu/abs/2012ApJ...744...44W} {744, 44}

\bibitem[\protect\citeauthoryear{{Wetzel}, {Hopkins}, {Kim},
  {Faucher-Gigu{\`e}re}, {Kere{\v s}}  \& {Quataert}}{{Wetzel}
  et~al.}{2016}]{2016ApJ...827L..23W}
{Wetzel} A.~R.,  {Hopkins} P.~F.,  {Kim} J.-h.,  {Faucher-Gigu{\`e}re} C.-A.,
  {Kere{\v s}} D.,   {Quataert} E.,  2016, \mn@doi [\apjl]
  {10.3847/2041-8205/827/2/L23}, \href
  {http://adsabs.harvard.edu/abs/2016ApJ...827L..23W} {827, L23}

\bibitem[\protect\citeauthoryear{{Whitaker}, {van Dokkum}, {Brammer}  \&
  {Franx}}{{Whitaker} et~al.}{2012}]{2012ApJ...754L..29W}
{Whitaker} K.~E.,  {van Dokkum} P.~G.,  {Brammer} G.,   {Franx} M.,  2012,
  \mn@doi [\apjl] {10.1088/2041-8205/754/2/L29}, \href
  {http://adsabs.harvard.edu/abs/2012ApJ...754L..29W} {754, L29}

\bibitem[\protect\citeauthoryear{{Whitaker} et~al.,}{{Whitaker}
  et~al.}{2014}]{2014ApJ...795..104W}
{Whitaker} K.~E.,  et~al., 2014, \mn@doi [\apj] {10.1088/0004-637X/795/2/104},
  \href {http://adsabs.harvard.edu/abs/2014ApJ...795..104W} {795, 104}

\bibitem[\protect\citeauthoryear{{White} \& {Frenk}}{{White} \&
  {Frenk}}{1991}]{1991ApJ...379...52W}
{White} S.~D.~M.,  {Frenk} C.~S.,  1991, \mn@doi [\apj] {10.1086/170483}, \href
  {http://adsabs.harvard.edu/abs/1991ApJ...379...52W} {379, 52}

\bibitem[\protect\citeauthoryear{{Zahid}, {Dima}, {Kudritzki}, {Kewley},
  {Geller}, {Hwang}, {Silverman}  \& {Kashino}}{{Zahid}
  et~al.}{2014}]{2014ApJ...791..130Z}
{Zahid} H.~J.,  {Dima} G.~I.,  {Kudritzki} R.-P.,  {Kewley} L.~J.,  {Geller}
  M.~J.,  {Hwang} H.~S.,  {Silverman} J.~D.,   {Kashino} D.,  2014, \mn@doi
  [\apj] {10.1088/0004-637X/791/2/130}, \href
  {http://adsabs.harvard.edu/abs/2014ApJ...791..130Z} {791, 130}

\makeatother
\end{thebibliography}
\bibliographystyle{mnras}

\end{document}